\documentclass[table]{ametsocV6.1}

\usepackage{amsmath,amsfonts,amssymb,bm,multicol}
\usepackage{mathptmx}
\usepackage{newtxtext}
\usepackage{newtxmath}

\DeclareMathOperator{\sech}{sech}

\title{Measuring a rogue? An investigation into an apparent giant wave} 

\authors{Adi Kurniawan,\aff{a}\correspondingauthor{Adi Kurniawan, adi.kurniawan@uwa.edu.au}
Paul H. Taylor,\aff{a} 
Jana Orszaghova,\aff{a}  
Hugh Wolgamot,\aff{a}
and Jeff Hansen\aff{b}
}

\affiliation{\aff{a}{Oceans Graduate School, The University of Western Australia, WA 6009, Australia}\\
\aff{b}{School of Earth Sciences, The University of Western Australia, WA 6009, Australia}
}

\abstract{An apparent giant wave event having a maximum trough-to-crest height of 21 metres and a maximum zero-upcrossing period of 27 seconds was recorded by a wave buoy at a nearshore location off the southwestern coast of Australia. 
It appears as a group of waves which are significantly larger both in height and period than the waves preceding and following them. 
This paper reports a multifaceted analysis into the plausibility of the event. 
We first examine the statistics of the event in relation to the rest of the record, where we look at quantities such as maximum-to-significant wave height ratios, ordered crest-trough statistics, and average wave profiles. 
We then investigate the kinematics of the buoy, where we look at the relationship between the horizontal and vertical displacements of the buoy, and also attempt to numerically reconstruct the giant event using Boussinesq and nonlinear shallow water equations. 
Additional analyses are performed on other sea states where at least one of the buoy's accelerometers reached its maximum limit.   
Our analysis reveals incompatibilities of the event with known behaviour of real waves, leading us to conclude that it was not a real wave event.
Wave events similar to the one reported in our study have been reported elsewhere and have sometimes been accepted as real occurrences. 
Our methods of forensically analysing the giant wave event should be potentially useful for identifying false rogue wave events in these cases.}

\begin{document}

\maketitle

\section{Introduction}

On the south coast of Western Australia, swells significantly larger than the waves around them, locally known as king waves, have frequently been reported. 
They are said to appear in a wave train without warning, and a number of fatalities, especially among recreational rock fishermen, have been attributed to them. 
However, apart from eye-witness accounts~\citep{Underwood2013,DIDENKULOVA2020105076}, there is a great lack of quantitative measurements to confirm how large these events really are.  Compared to rogue waves that occur in deep water, coastal rogue waves, or rogue waves that occur in relatively shallow water, are not as well studied. 
However, several recent studies have reported evidences in support of the increasing probability of rogue waves as waves propagate from deep to shallow water~\citep{doi:10.1063/1.4748346,doi:10.1063/1.4847035,zhang2019statistics}.

As part of a greater effort to characterise the wave climate off the southern coast of Australia, a wave buoy was deployed off Torbay, about 12 km southwest of Albany, Western Australia.  
The wave buoy is a Directional Waverider 4 (DWR4), {a stabilised platform accelerometer-based 
wave buoy}~\citep{DWR4manual},
moored at 
35$^\circ$04'15.5''S 117$^\circ$46'29.9''E.
The water depth at the location is approximately 30 m and the distance from shore is approximately 1 km.

\begin{figure*}
\centering
\includegraphics[width=39pc]{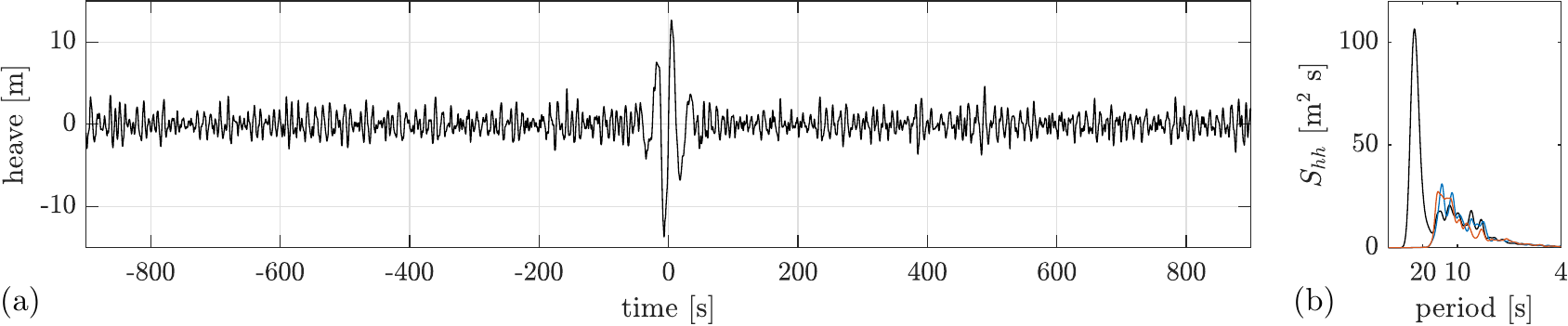}
\caption{(a) Time history of the vertical displacement of the buoy, 
centred at the giant wave group. (b) The corresponding spectrum {(black) and the spectra for 30-min intervals before (blue) and after (red) the wave group}.}
\label{heave_giant}
\end{figure*}

Out of the wave record so far collected by the buoy, an interesting event was recorded by the buoy on 19 September 2019 (Fig.~\ref{heave_giant}). 
It appears as a group of waves which are significantly larger both in height and period than the waves preceding and following them.
The buoy continued to function after the event, suggesting that there was no damage to the buoy or its mooring 
caused by 
these waves. 
Although the event appears to be a candidate for a rogue wave event, the wave group appears to be disproportionately large relative to the background waves as well as to the water depth (30 m), giving doubts to its veracity. 


{Surface-following buoys are by far the most popular means for field wave measurements.}  
An accelerometer-based buoy records its accelerations, which are then doubly integrated on board to give the displacement time series. 
It has been pointed out that realistically shaped but unrealistically large waves can be caused by spikes in the acceleration time series~\citep{soton79448,Cahill2013}.
While this could well be the source of the apparent giant wave group in our record, we note that the displacement time series reported in the above literature contain clear contamination (in the form of unrealistically looking high sharp spikes) at about 100 s prior to the large events, whereas we do not observe such contamination in our time series. 

{Before immediately dismissing this event as spurious, therefore,} 
in this paper we subject the giant wave event to a detailed investigation. 
Given the stochastic nature of ocean waves, distinguishing false rogue wave events from the real ones is not a straightforward task, precisely because rogue waves by definition are waves that are exceptionally larger than expected for the given sea state~\citep{Dysthe2008}. 
Having more than one independent source of measurements is helpful to validate the event, but often these are not available. 
In this case, the plausibility of the event will need to be deduced through other means. 
In this paper, {this is done by examining} the statistics {(Section~\ref{secstatistics})} as well as the kinematics {(Section~\ref{seckinematics})} of {the event based on} the recorded displacements of the buoy{, and comparing them against the background of a one-year long record}. 
Comparisons are also made with other {potentially spurious} events in the record{, where at least one of the accelerometers in the buoy exceeded its limit} {(Section~\ref{sec:accelerometers})}. 
{Additional means to analyse the event are discussed in Section~\ref{secaddanalysis}.}  
{Each of the analyses described in this paper is suggestive on its own, but considering all the analyses together we believe rules out the physical plausibility of this giant wave event.
A table summarising the various tests presented in this paper is given in Section~\ref{sectests} together with some suggested thresholds, which potentially could be applied on any dataset to test the plausibility of contentious rogue waves.}

\section{Statistics} \label{secstatistics}


The buoy records its north, west, and vertical displacements at a sampling rate of 5.12 Hz and outputs them at a rate of 2.56 Hz. 
In addition, every 30 minutes, it outputs bulk statistics, which are processed on-board. 
The buoy on-board data processing applies a bandpass filter with cutoff periods of 1 s and 30 s.
The GPS location of the buoy (latitude and longitude) is  output every 10 minutes with an accuracy of approximately 5 m. 



A one-year long full displacement record spanning from 1 October 2018 to 1 October 2019 is analysed. 
This is the data stored on board the buoy and not transmitted.
The record is broken into 30 minutes long intervals, which we define here as sea states.
Within the one-year record, there are in total six sea states with missing displacement data points, which happened simultaneously for the heave, north, and west displacements.
The number of missing data points ranges from one to four consecutive points, or equivalent to a duration of 1.95 s at the most.
We fill the missing points using spline interpolation.
Missing data points also occurred within the giant wave group and will be discussed in more detail in Section~\ref{sec:accelerometers}.

\subsection{Wave heights and periods}

Zero-upcrossing analysis is performed on the raw vertical displacement of the buoy to obtain the wave heights and wave periods of the individual waves.
From here the maximum wave height and period in the sea state, $H_\mathrm{max}$ and $T_\mathrm{max}$, {the maximum crest heights $\eta_{c,\mathrm{max}}$,} as well as the mean upcrossing period $T_z$ {are obtained}.  
The significant wave height $H_s$ is computed as $4\sigma$, where $\sigma$ is the standard deviation of the vertical displacement. 
These independently calculated values agree closely with corresponding values processed on-board by the buoy and output in the bulk statistics data files. 
The only exception is the $H_\mathrm{max}$ and $T_\mathrm{max}$ of the sea state in which the giant wave event occurred, which were probably excluded due to the quality control implemented in the buoy.
{We note from the manufacturer that} the buoy flags events where the pitch or roll angles
exceed 89$^\circ$ or the accelerations exceed 1 g (standard {acceleration due to} gravity), and excludes segments containing such exceptions from the spectral calculations. 
{However, ``the occurrence of a flag does not imply an error, nor does the absence of a flag warrant a correct measurement. A flag is merely indicative of a sensor having got near the limits of its range''}~\citep{DWRcomparison}.

\begin{figure}
\centering
\includegraphics[width=27pc]{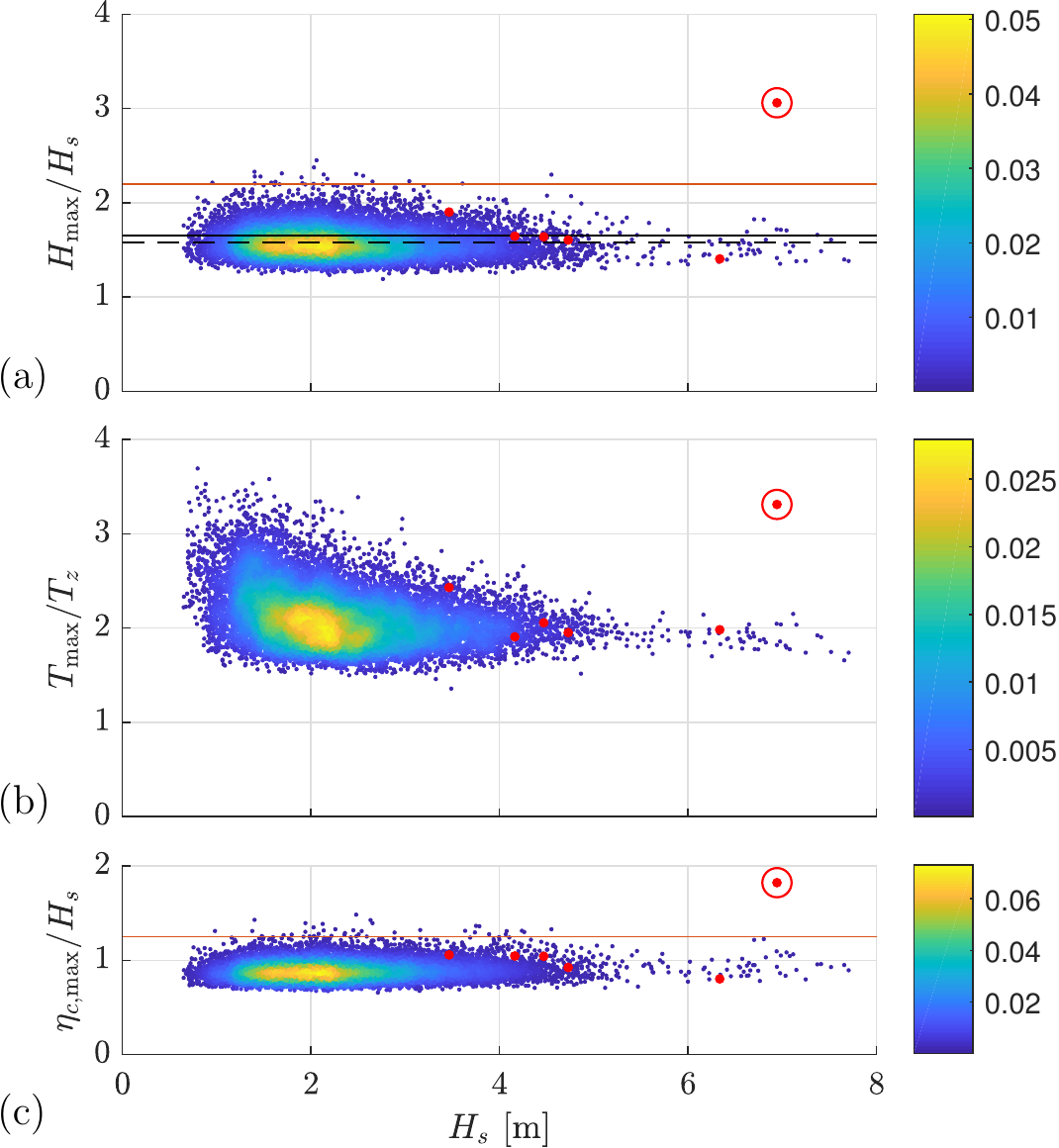}
\caption{(a) $H_\mathrm{max}/H_s$, (b) $T_\mathrm{max}/T_z$, and {(c) $\eta_{c,\mathrm{max}}/H_s$} ratios of all 30-minute sea states in the one-year record, plotted against $H_s$. 
{Color indicates point density.}
The sea states in which at least one of the accelerometers reached its maximum limit are marked in {red}.
The sea state in which the giant wave event occurred is {enclosed with a red circle}. 
In (a), solid {black} line indicates the mean of the most probable maximum values calculated according to the Rayleigh distribution, while 
dashed line indicates the mean of all the observations. {The solid red line indicates the $H_\mathrm{max} = 2.2 H_s$  threshold. In (c), solid red line indicates the $\eta_{c,\mathrm{max}} = 1.25 H_s$ threshold.}  
}
\label{HTratio}
\end{figure}

Based on the zero-upcrossing analysis, the maximum wave height $H_\mathrm{max}$ of the giant wave group was 21.25 m (the zero-downcrossing maximum wave height is greater, 26.37 m), while {the maximum crest height $\eta_{c,\mathrm{max}}$ was 12.66 m and} the maximum wave period $T_\mathrm{max}$ was 26.84 s. 
These were also the largest wave height and period in the entire one-year record.
For the sea state in which the giant wave event occurred, we have $H_s = 6.94$ m and $T_z = 8.10$ s, giving $H_\mathrm{max}/H_s = 3.06${, $\eta_{c,\mathrm{max}}/H_s = 1.82$,} and $T_\mathrm{max}/T_z = 3.31$. 
Fig.~\ref{HTratio} shows how these values compare with those of the rest of the sea states in the year. 
{In each of the plots, the giant wave appears as an outlier.}

\begin{figure}
\centering
\includegraphics[width=27pc]{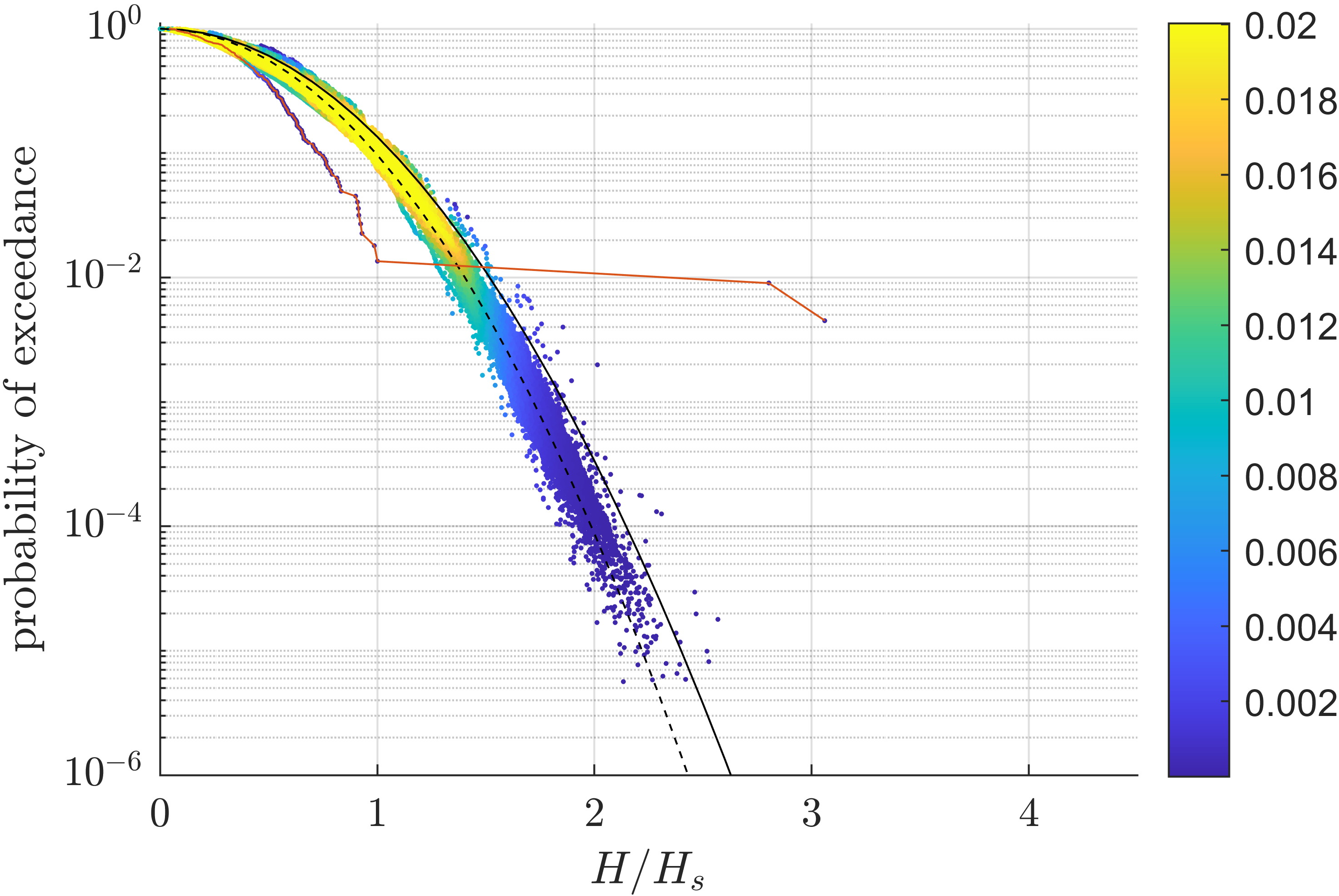}
\caption{{Distribution of normalised zero-upcrossing wave heights in the one-year record, with colour indicating point density.
Sea states are binned according to $H_s$ and $T_z$, where the bin size is 0.5 m by 0.5 s.   
Solid black line is the Rayleigh distribution (with the RMS amplitude taken as $\sqrt{2}\sigma$), whereas the dashed line is 
the form suggested by~\citet{Longuet-Higgins1980} (with the RMS amplitude taken as $0.925\sqrt{2}\sigma$).
The wave heights in the sea state in which the giant wave event occurred are connected by a red line.} 
}
\label{Hdist}
\end{figure}

\begin{figure}
\centering
\includegraphics[width=27pc]{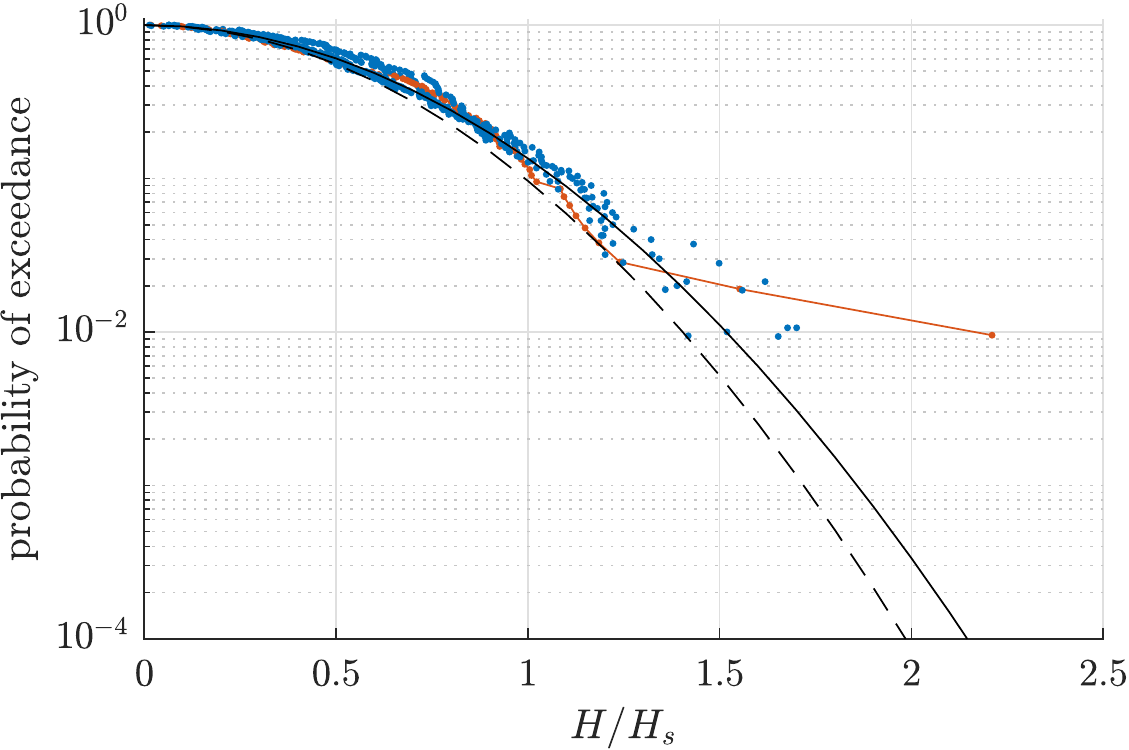}
\caption{{Distribution of normalised zero-upcrossing wave heights for six 20-min records around the Draupner wave. 
Solid black line is the Rayleigh distribution (with the RMS amplitude taken as $\sqrt{2}\sigma$), whereas the dashed line is 
the form suggested by~\citet{Longuet-Higgins1980} (with the RMS amplitude taken as $0.925\sqrt{2}\sigma$).
The wave heights in the sea state in which the Draupner wave occurred are connected by a red line.} 
}
\label{HdistDraupner}
\end{figure}

The value of $H_\mathrm{max}/H_s = 3.06$ is significantly higher than the most probable maximum wave height ratio expected within the 30-min sea state, which, assuming Rayleigh distribution, is found to be $1.64$ according to
\begin{equation}
    \frac{H_\mathrm{max}}{H_s} = \sqrt{\frac{1}{2} \log \left(\frac{t}{T_z}\right)},
\end{equation}
with $t$ being the duration of the sea state.
{The mean of the theoretical most probable maximum {wave heights} from all sea states is close to the mean of the recorded $H_\mathrm{max}$ values, as shown in Fig.~\ref{HTratio}a.
This suggests that the distribution of the wave heights in each sea state is otherwise consistent with
a Rayleigh model.}
{To test this further, we plot the distribution of the normalised wave heights in Fig.~\ref{Hdist}.
The distributions, apart from the sea state of the giant event, are seen to be 
consistent with
Rayleigh.}
{%
Two forms of the Rayleigh model are plotted in this figure: the widely-used form with the RMS amplitude taken as $\sqrt{2}\sigma$ and a form with the RMS amplitude taken as $0.925\sqrt{2}\sigma$, as suggested by~\citet{Longuet-Higgins1980}. 
The latter is seen to give a better fit to the observations.
We note in passing that several other forms exist, such as described by~\citet{Forristall1984,TAYFUN20071631}. 
Regardless of which form we use,  
the giant wave noticeably skews the distribution of  wave heights in the same sea state.
This bias is due to an increased $H_s$ of approximately 1.5 times that of the background waves, owing to the presence of the giant wave.}
{For comparison, the normalised wave heights for six 20-min intervals around the well-known Draupner wave~\citep{haver2004possible} are plotted in Fig.~\ref{HdistDraupner}.
We remark that the Draupner wave  was recorded by a downward-pointing laser, i.e.~an Eulerian sensor instead of a Lagrangian sensor such as in the present case.
Each of the six distributions is 
consistent with
Rayleigh.
The Draupner wave skews the distribution of the wave heights in the same 20-min interval, but only at the tail.} 

The probability of $H_\mathrm{max}$ exceeding a certain value $h$ in a {sea state} of $t$ duration is by Rayleigh distribution given as
\begin{equation}
    P(H_\mathrm{max} \geq h) = 1 - \left\{1 - \exp \left[-2 \left(\frac{h}{H_s}\right)^2\right]\right\}^{\frac{t}{T_z}}.
\end{equation}
From this, the probability of $H_\mathrm{max} \geq 3.06 H_s$ in a 30-min record is found to be 
{$1.60 \times 10^{-6}$. 
In order for a giant wave of this size to be the most probable maximum wave in a sea state of this severity, the sea state would have to have lasted 36 years.}
%
As a comparison, the second highest $H_\mathrm{max}/H_s$ ratio in the year was 2.45.
The probability of $H_\mathrm{max} \geq 2.45 H_s$ {in a record consisting of approximately 200 waves} is in the order of $10^{-3}$.
{The probability of $H_\mathrm{max} \geq 2.15 H_s$,
such as that of the Draupner wave ($T_z = 11.28$ s), is an order of magnitude lower. 
The sea state would have to have lasted 1.31 days for a wave of this size to be the most probable maximum wave in a sea state of this severity.
Even in terms of its $\eta_{c,\mathrm{max}}/H_s$ ratio, which at a value of 1.55 is  impressive, the corresponding return period of the Draupner is only approximately 2.3 months~\citep{Dysthe2008}.}


If we consider a 30-min record centred at the middle zero-upcrossing of the giant wave {group (see Fig.~\ref{heave_giant})}, the maximum wave height in this sea state was 3.02 {times} larger than the largest wave height found before and after the group, while the maximum wave period was 1.70 times longer than the longest period outside the group.
Within the one-year record, the largest wave height outside the giant wave group was 12.40 m. 
Thus the maximum wave height in the giant wave group is at least 1.71 times as large as the rest of the waves outside the group. 
On the other hand, waves with periods close to the maximum wave period in the giant wave group occurred more frequently. 
The longest wave period outside the giant wave group was 25.44 s.
This is close to 26.84 s, the maximum wave period in the group, and is in fact longer than the second longest wave period within the group.
However, {while the observed period of the giant wave is not so rare, in the $(H_s, T_\mathrm{max}/T_z)$ space (Fig.~\ref{HTratio}b), the giant wave appears as a clear outlier.}
{Given that there is not yet a consensus on whether} rogue wave events {originate} from the same distribution as that of the background waves~{\citep{fedele2016real}} or originate from an entirely different process~{\citep{haver2000rare}}, this {arguably} does not necessarily disprove this event. 
{However, 
%
it is informative to compare the characteristics of our giant wave with those of other rogue waves that are known to be real.
Table~\ref{extremewaves} compares our giant wave with the much studied Draupner and Andrea waves~\citep{Magnusson2013Andrea}. 
The normalised wave height and crest height of our giant wave are higher than those of Draupner and Andrea, but the most striking difference is in the normalised period, which is more than twice of Draupner's or Andrea's.
Note that there is some ambiguity on the definition of wave height and period in the literature, such as whether it is taken between zero upcrossings or zero downcrossings, but the different definition only slightly changes the normalised values given in the table.}

\begin{table*}
{%
\caption{{Characteristics of the giant wave in the present paper compared to those of Draupner and Andrea.}}
\label{extremewaves}
\begin{center}
\begin{tabular}{cccccl}  
\topline
Event    & Date  & $H/H_s$ & $\eta_c/H_s$ & $T/T_z$ & Sensor \\
\midline
Present paper& 19 Sep 2019    & 3.06 & 1.82 & 3.31 & Accelerometer-based buoy \\
Draupner     & 1 Jan 1995     & 2.15 & 1.55 & 1.42 & Laser \\
Andrea     & 9 Nov 2007     & 2.49 & 1.63 & 1.33 & Laser \\
\botline
\end{tabular}
\end{center}
}
\end{table*}

It is worth noting that%
~\citet{nhess-18-729-2018} reported $H_\mathrm{max} = 29.96$ m and $H_s = 7.15$ m from a 30-minute record of a Waverider buoy on 8 December 2014 near Killard, Ireland. 
This gives $H_\mathrm{max}/H_s = 4.19${, higher than the ratio of our giant wave}.
The record actually shows that the preceding wave was higher, possibly greater than 40 m, but the peak and trough were both clipped due to the sensors on the buoy only able to record a displacement up to $\pm 20.48$ m.
They also reported $H_\mathrm{max}/H_s = 18.5\text{ m}/4.41\text{ m} = 4.20$
from another Waverider buoy off Belmullet, Ireland, on 16 October 2017 during Storm Ophelia.
The Killard giant wave is similar in characteristics to our giant wave, 
{in that both have remarkably high $H_\mathrm{max}/H_s$ and $T_\mathrm{max}/T_z$ ratios,}
suggesting that if our giant wave is spurious, then this one probably is as well. 
No time history plot was presented for the Belmullet wave. 

\begin{figure}
\centering
\includegraphics[width=27pc]{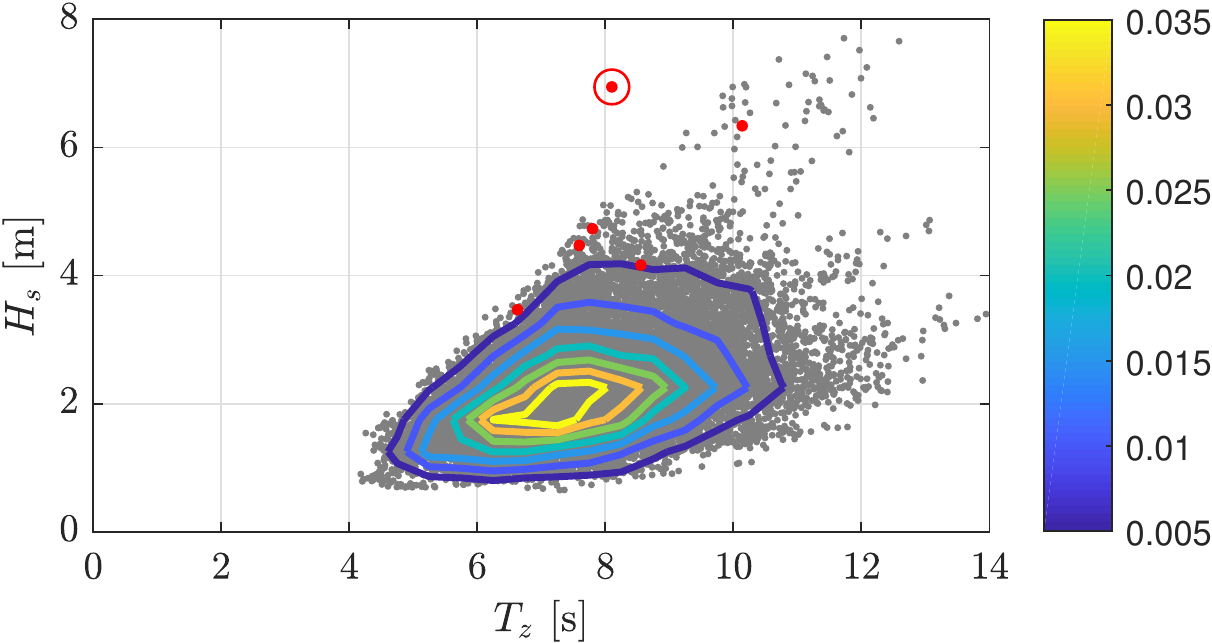}
\caption{Scatter plot of $H_s$ and $T_z$ of all 30-minute sea states in the one-year record.
Contour lines indicate the probability of occurrence.
The sea states in which at least one of the accelerometers reached its maximum limit are marked in {red}.
The sea state in which the giant wave event occurred is {enclosed with a red circle}. }
\label{scatter_plot}
\end{figure}

A scatter plot of the significant wave height $H_s$ and mean period $T_z$ of all the sea states in the entire one-year duration is shown in Fig.~\ref{scatter_plot}.
The sea state in which the giant wave event occurred appears to be {an outlier in terms of its steepness.}
The mean period $T_z$ of the sea states ranges from 4 s to 14 s, corresponding to $k(T_z)d$ value ($k$ and $d$ being the wavenumber and the water depth, respectively) between 7.5 and 0.9, 
such that the average waves at this water depth would be categorised as intermediate to deep water waves.
However, for the giant wave, the value of $k(T_\mathrm{max})d$ is 0.42, indicating a
shallow water wave. 

\subsection{Ordered crests and troughs}

A wave buoy spends more time at the crests, where it is moving forward with the waves, than at the troughs, where it is moving backward. 
For this reason, the Lagrangian mean {free-}surface elevation, i.e.~the mean {free-surface elevation} as seen from a reference frame moving with the buoy, is higher than the Eulerian mean, i.e.~the mean as seen from a fixed reference. 
In the deep-water limit and up to second order, this displacement of the mean elevation is equal to the amount that the crest is raised~\citep{srokosz_longuet-higgins_1986}.
However, an accelerometer-based wave buoy does not notice this change in the mean elevation, 
and thus
the increase in the crest elevation is 
not captured~{\citep[see, e.g.,][]{Forristall2000}}.
{Moreover,} in practice, the buoy's filter cutoff period of 30 s effectively removes a large proportion of the subharmonic components that may be present~\citep{10.1175/JPO-D-19-0228.1}, which is in the order of the wave group period.
In moderately shallow water, however, {the situation is different.}
The second-order contribution to the crest elevation and the Lagrangian mean do not completely cancel out and, as a result, some of the nonlinearity should be captured by the buoy~{\citep{10.1175/JPO-D-15-0129.1}}.

\begin{figure*}
\centering
\includegraphics[width=33pc]{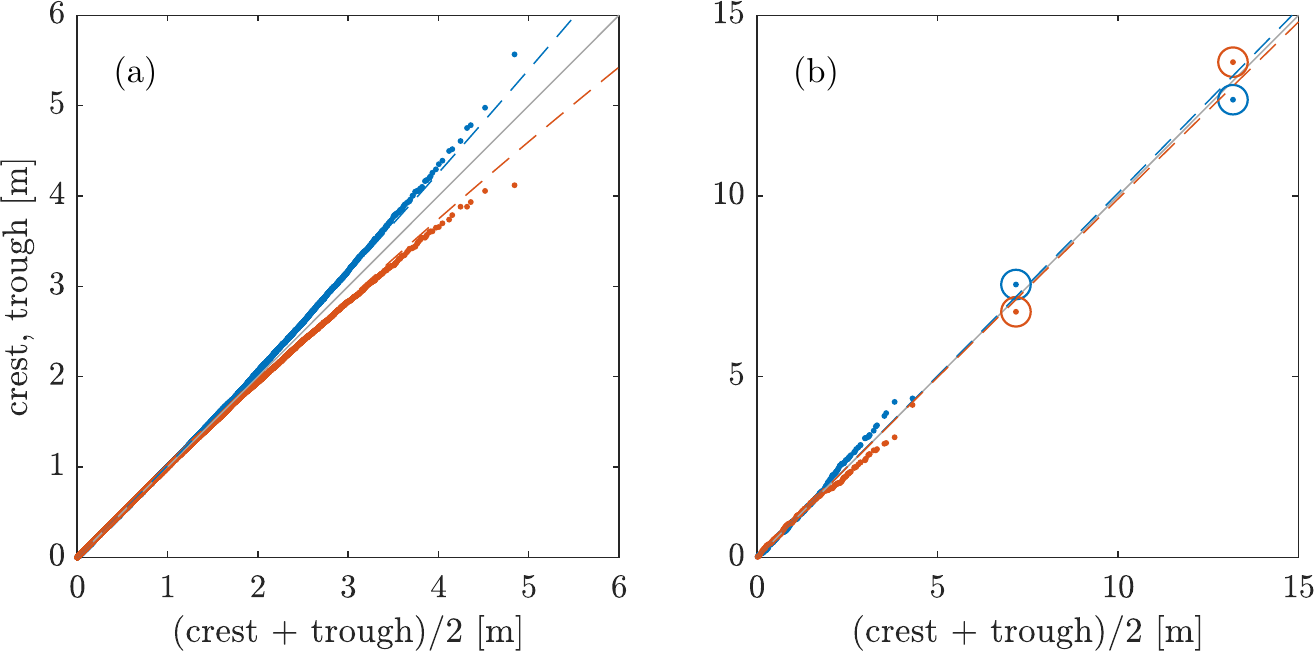}
\caption{{(a) Ordered crest elevations (blue dots) and trough depressions (red dots) versus the mean of the two, and the least-squares fit second-order models according to Eq.~\eqref{eqctfit} (dashed lines), 
for (a) a typical sea state ($H_s = 4.19$ m, $T_z = 8.73$ s), and
(b) the sea state with the giant wave. 
Circles mark the two highest crests and the two lowest troughs.}%
}
\label{cresttrough}
\end{figure*}

This is evident {from} Fig.~\ref{cresttrough}a, which shows an example of the ordered crest elevations and the ordered trough depressions of all the waves in a {typical, steep} sea state {with $T_z$ close to that of the sea state of the giant wave}. 
The ordered crest elevations are higher, while the ordered trough depressions are lower, than the mean of the two.  
Assuming that subharmonic components are completely removed, we may fit the following second-order model: 
\begin{equation}
    y = \begin{cases} x (1 + cx) & \text{for crest elevations}  \\
    x (1 - cx) & \text{for trough depressions}. \end{cases} \label{eqctfit}
\end{equation} 
The least-squares fit, as shown by the dashed lines in Fig.~\ref{cresttrough}a, approximates the distributions of the crests and troughs quite well, except at the 
higher ends, where terms beyond second order become significant.
{For comparison, the corresponding plot for the sea state of the giant wave is shown in Fig.~\ref{cresttrough}b.
The two highest crests and the two lowest troughs do not follow the trend, with the largest crest elevation being lower and the largest through depression being higher than the mean of the two. 
This results in $c = 8.1\times 10^{-4}$, much smaller than expected for this sea state.
For Fig.~\ref{cresttrough}a, $c = 0.016$.}

Neglecting directional spread, we may estimate the second-order Lagrangian superharmonic component using the expression derived by~\citet{WHITTAKER2016253}, which is valid for unidirectional regular waves in finite water depth. 
According {to this expression, the coefficient $c$ in the second-order model~\eqref{eqctfit} is given as}
\begin{multline}
    c = \frac{3}{2+C} \frac{1+2D}{2(1-D)} k \coth(kd), \\ \text{where } C = \cosh(2kd) \text{ and } D = \sech(2kd) . \label{eqcunireg}
\end{multline} 

\begin{figure}
\centering
\includegraphics[width=27pc]{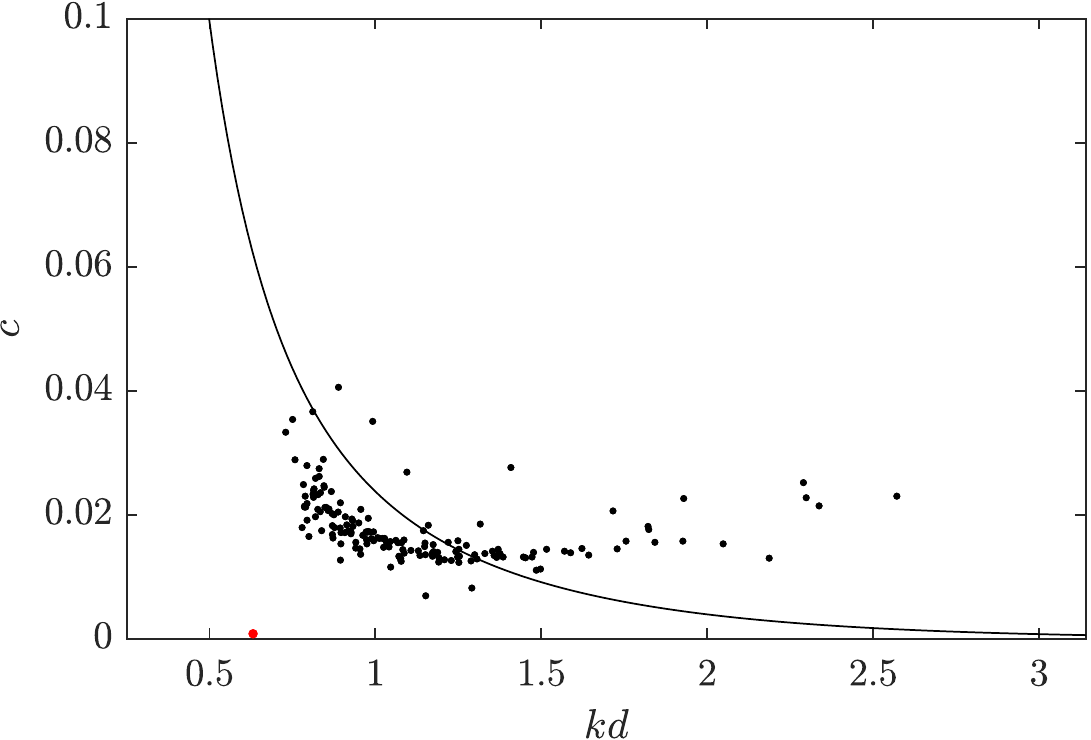}
\caption{Coefficients $c$ obtained by least-squares fitting of Eq.~\eqref{eqctfit} to the ordered crest elevations and trough depressions from each sea state (black dots), and the theoretical $c$ according to Eq.~\eqref{eqcunireg}, which is valid for unidirectional regular waves (solid line).
The coefficient $c$ for the sea state containing the giant wave is shown in red. 
The effective wavenumber is taken as the wavenumber at the energy period $T_e$.%
}
\label{coeff}
\end{figure}

Figure~\ref{coeff} compares this theoretical estimate of $c$ {(obtained from~\eqref{eqcunireg})} with the least-squares coefficients $c$ {(obtained from~\eqref{eqctfit}), for} all sea states in the record. 
{In the calculations, the sea states are binned according to the $H_s$ and $T_z$ values, using 0.5 m by 0.5 s as the bin size.}
{Binning is important to obtain an accurate fit, thus minimising the spread of the least-squares $c$ values.}
{The coefficients are plotted} as a function of $k(T_e) d$, where $T_e$ is the {mean} energy period {of all the sea states in the bin}. 
{The energy period is defined as $T_e = m_{-1}/m_0$, where $m_n$ is the $n$th spectral moment~\citep{TUCKER1993459}.
Since~\eqref{eqcunireg} is for regular waves, a representative wave period is needed to calculate the wavenumber $k$.
Here we have used $T_e$, but  $T_z$ can also be used, leading in principle to the same conclusion, i.e.~the sea state in which the giant wave resides appears as an outlier.
The contrast is however clearer when $T_e$ is used.}
The agreement {of the model with the other sea states is otherwise}  reasonably good considering the approximate nature of the analysis {(note that the theoretical expression~\eqref{eqcunireg} is for unidirectional regular waves, as opposed to directionally spread irregular waves)}.
{To be more accurate,} one should account for bandwidth by using the full second-order theory rather than regular wave forms, and for directional spreading {rather than assuming unidirectional propagation}, but it is reassuring that simple theory can account for the crest-trough asymmetry reasonably well, suggesting that the measured data---apart from the giant wave---are reliable. 

\subsection{Average wave profiles}

Assuming that the heave displacement of the buoy is a Gaussian process with a zero mean, we can obtain the expected heave displacement around a point in time, given the displacement and its derivatives at that particular time. 
The expression, using information up to the third derivative, has the following form:%
\begin{equation}
    \xi(t) = \xi(0) A(t) + \dot{\xi}(0) B(t) + \ddot{\xi}(0) C(t) + \dddot{\xi}(0) D(t), \label{expshape}
\end{equation}
where $\xi(0)$, $\dot{\xi}(0)$, $\ddot{\xi}(0)$, and $\dddot{\xi}(0)$ are the displacement and its time derivatives at the conditioning point, defined as $t = 0$%
~\citep[see][]{10.2307/2240325,10.2307/1403443}. 
Similar expressions applicable for more specific cases such as the expected profile  around a most probable maximum~\citep{tromans1991new} and around an arbitrary maximum or minimum~\citep{10.2307/2240325,10.1115/1.2829043} have been widely used. 
The functions $A(t)$, $B(t)$, $C(t)$, and $D(t)$ in Eq.~\eqref{expshape} depend on the autocorrelation function $R_{hh}(t)$ of the record, which in turn is related to the {heave} displacement spectrum $S_{hh}(f)$ by Fourier transform:%
\begin{equation}
    R_{hh}(t) = \int_0^\infty S_{hh}(f) \cos(2\pi f t) \,\mathrm{d}f. 
\end{equation}
Specifically, $A(t)$, $B(t)$, $C(t)$, and $D(t)$ are the columns of $\bm{\mathsf{R}} \bm{\mathsf{L}}^{-1}$, where
\begin{equation}
    \bm{\mathsf{R}} = \begin{pmatrix}
    r(t) & -\dot{r}(t) & \ddot{r}(t) & -\dddot{r}(t)
    \end{pmatrix} \label{eqR} 
\end{equation} and
\begin{equation}
    \bm{\mathsf{L}} = \begin{pmatrix}
    1 & 0 & -\lambda_2 & 0 \\
    0 & \lambda_2 & 0 & -\lambda_4 \\
    -\lambda_2 & 0 & \lambda_4 & 0 \\
    0 & -\lambda_4 & 0 & \lambda_6
    \end{pmatrix}. \label{eqL}
\end{equation}
Here, $\lambda_n$ and $r(t)$ are the $n$-th spectral moment $m_n = \int_0^\infty{\omega^n S_{hh}(\omega)}\, \mathrm{d}\omega$ and the autocorrelation function $R_{hh}(t)$ both normalised by the zeroth moment $m_0$.

\begin{figure*}
\centering
\includegraphics[width=39pc]{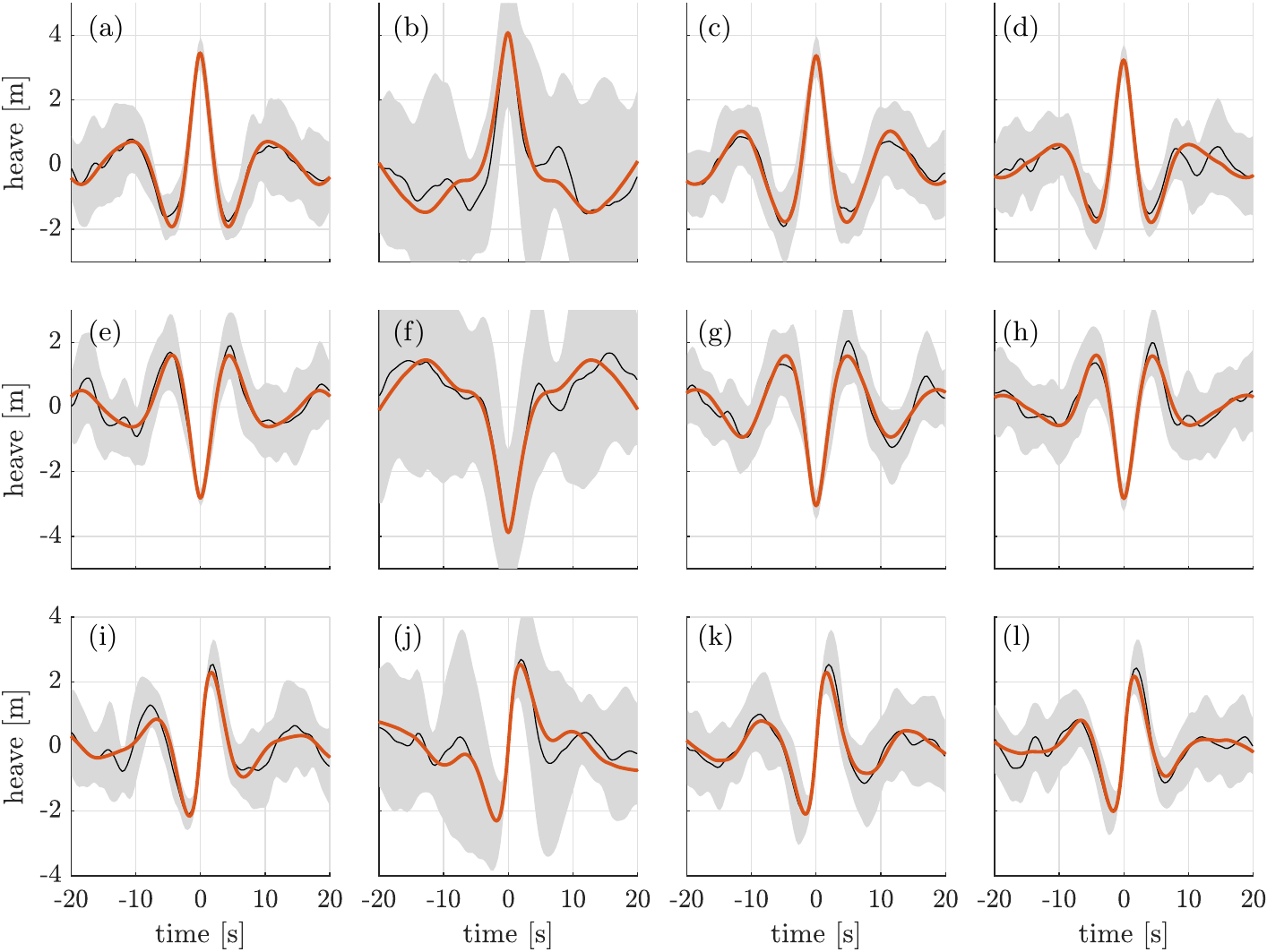}
\caption{{(a--d) Measured average profile around the highest 20 crests in a 30-minute record (black) and the theoretical expected profile (red), for (a) a record ending just before the giant wave group, (b) a record centred at the giant wave, (c) a record starting just after the giant wave group, and (d) a record centred at the giant wave but with the giant wave group ($-50 < t < 50$ s)  ramped down to zero. Grey band indicates $\pm \sigma_s$, where $\sigma_s$ is the sample standard deviation. (e--h) The same for the lowest 20 troughs, and (i--l) for the steepest 20 slopes.}%
}
\label{avgshape}
\end{figure*}

This allows us to verify if the measured average wave profile around a point of similar characteristics behaves as expected from the underlying spectrum. 
{The expected profile from a record centred at the giant wave group is analysed by comparing it to the expected profiles from equally long records ending just before the group and starting just after the group, as well as a record centred at the group but with the giant wave ramped down to zero.} 
Such verification is shown in Figure~\ref{avgshape},  which shows the average heave displacement of the buoy around the highest 20 crests, lowest 20 troughs, and steepest 20 slopes in the 30-minute records.
{The measured average has been obtained by averaging the displacement as measured by the buoy, i.e.~without linearisation.} 
{The average profiles from the record containing the giant wave (Fig.~\ref{avgshape}b, f, j) appear to be different from the average profiles taken from records without the giant wave (where all have similar profiles close to the expected theoretical ones and relatively small standard deviation around the mean). 
The crest height and trough depression are larger for the average profiles from the sea state containing the giant wave (b and f). 
Away from the conditioning point, the expected profile of the highest crests (lowest troughs) is not immediately followed by troughs (crests) as in the other records without the giant wave. 
Similarly, for the steepest slopes, the profile away from the conditioning point is different from the profile of the other records.
This points to the fact that the giant wave modifies the underlying spectrum. 
Moreover, the poorer agreement between the measured average and the theoretical average, as well as the larger standard deviation when the giant wave is included suggest that the giant wave does not behave as expected from the underlying spectrum.} 

\begin{figure}
\centering
\includegraphics[width=27pc]{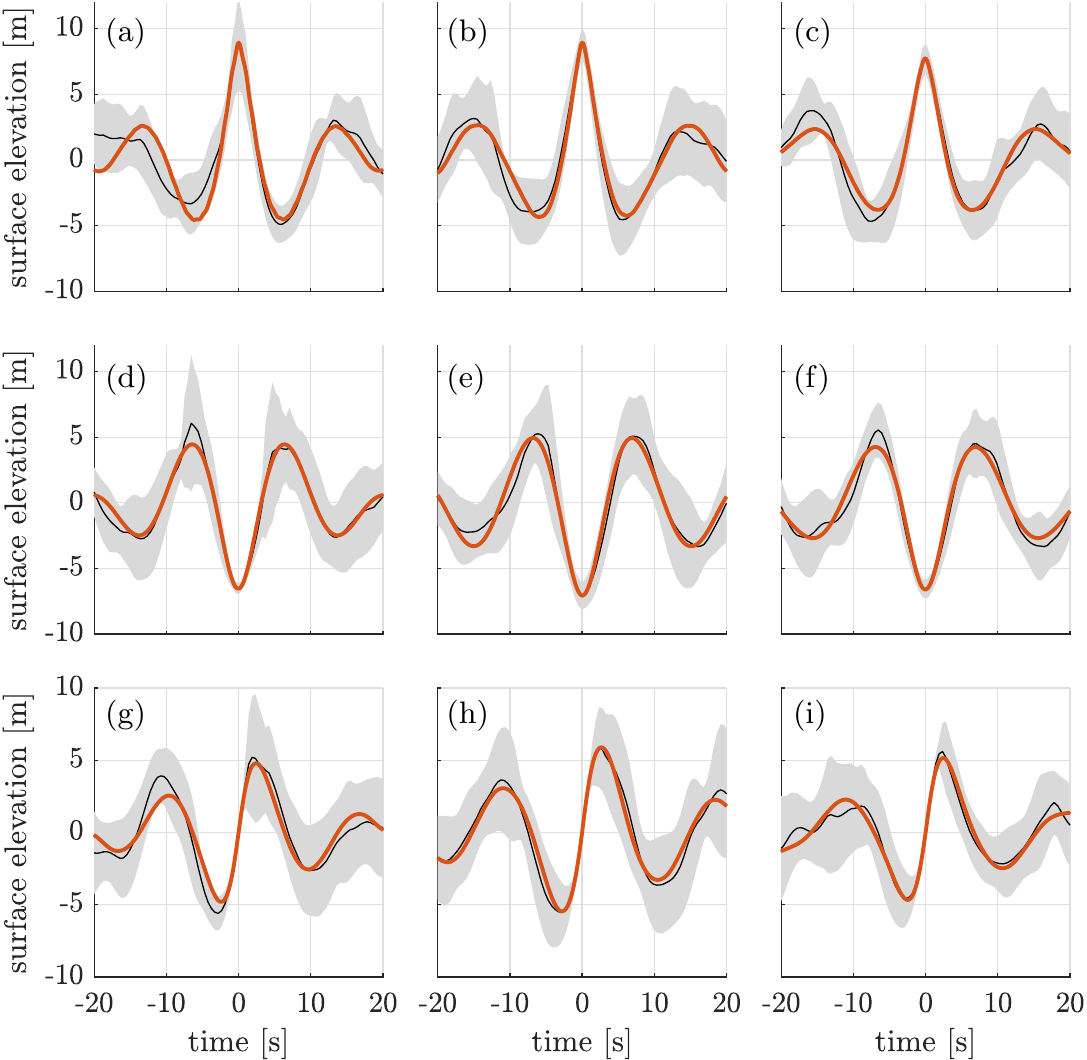}
\caption{{(a--c) Measured average profile around the highest 10 crests in a 20-minute record (black) and the theoretical expected profile (red), for (a) a record with the Draupner wave, and (b--c) two records within the same 3-hour interval. Grey band indicates $\pm \sigma_s$, where $\sigma_s$ is the sample standard deviation. (d--f) The same for the lowest 10 troughs, and (g--i) for the steepest 10 slopes.}%
}
\label{avgshapedraupner}
\end{figure}

{For comparison, the expected profiles for the Draupner wave are shown in Fig.~\ref{avgshapedraupner}.
We compare the average profiles from the 20-min interval containing the Draupner wave and another two 20-min intervals within the same 3-hour interval.
Due to the shorter interval and longer $T_z$---thus fewer waves in the sea state---we average 10 instead of 20 profiles. 
The average measured profiles and the theoretical profiles are similar across the three intervals, with comparable standard deviation, suggesting that the Draupner wave behaves as expected from the underlying spectrum.
Furthermore, although the sea state of the Draupner ($H_s = 11.92$ m and $T_z = 11.28$ s) has different $H_s$ and $T_z$ than the sea states around our giant wave event ($H_s \approx 4.7$ m and $T_z \approx 8.0$ s), their average profiles have the same characteristics, in that the highest (lowest) crest (trough) is immediately followed by a trough (crest) on either side.
}

\section{Kinematics} \label{seckinematics}

\subsection{Wave height to water depth ratio}

For the giant wave, the ratio of the maximum wave height to the water depth of 30m is {0.71 (zero-upcrossing) and 0.88 (zero-downcrossing).} 
The limiting value (due to wave breaking) is known to range
from 0.55 for a horizontal or gently sloping seabed to more than 1 for a seabed with a steeper gradient such as 1:8~\citep{herterich_dias_2019}. 
The seabed profile in our case has an approximate gradient of {1:60 at the buoy location (approximately 1000 m offshore){; cf.~Fig.~\ref{Bouss}b}. 
This increases almost linearly to 1:30 at about 500 m offshore; 
while further offshore, 
the gradient is approximately {1:90} (2000 m offshore).
These values are based on bathymetric data obtained from a 2018 LiDAR survey at and around the buoy's location.}
{While our observed $H_\mathrm{max}/d$ values may seem rather high, they are in fact in agreement with those predicted using empirical formulae for $H_\mathrm{max}/d$ on sloping beds suggested in~\citet{nelson1987design,Allsop1998}.}

\subsection{Horizontal to vertical displacement amplitude ratio {and relative phase}} 

We may resolve the east and north displacements of the buoy into the mean direction of the waves, which we define here as the surge direction.
The mean direction {is} obtained from the lowest-order Fourier coefficients of the directional distribution $G(\theta, f)$, which are related to the heave, north, and east displacement spectra of the buoy
as well as the cross-spectra%
~\citep[see, e.g.,][]{LygreKrogstad1986}. 

\begin{figure*}
\centering
\includegraphics[width=39pc]{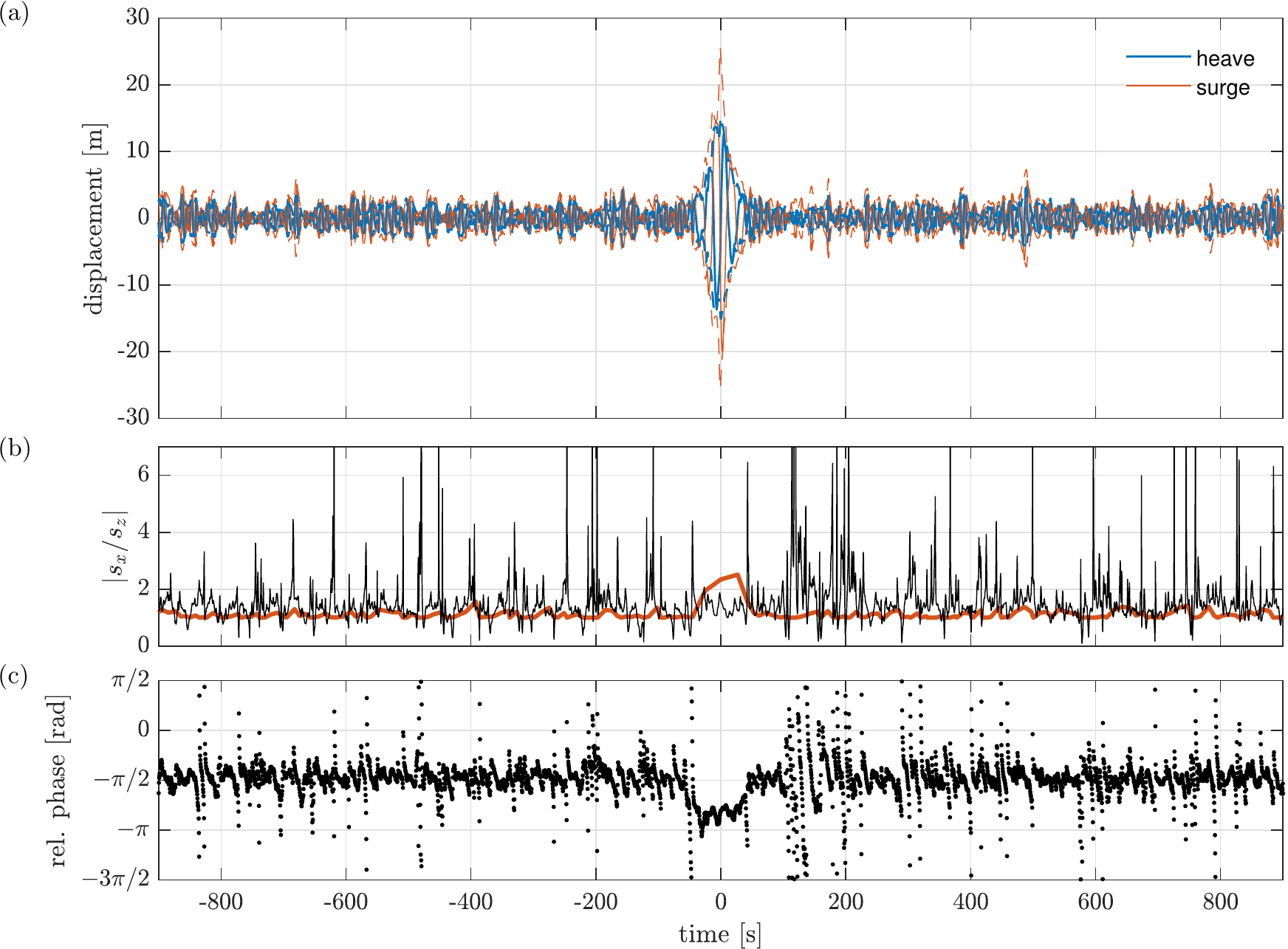}
\caption{(a) Heave and surge displacements of the buoy (solid lines) and their envelopes (dashed lines) {in the sea state centred} around the giant wave event. 
(b) Ratio of the surge to heave displacement envelopes {(black)} and the theoretical ratio $\coth(kd)$  according to linear theory {(red)}, with $k$ calculated from the zero-upcrossing wave periods.
{(c) Instantaneous phase of surge displacement relative to heave.}
}
\label{xzratio}
\end{figure*}

The ratio of the horizontal to vertical amplitudes of the buoy displacement is given as $\coth(kd)$ according to linear theory, where $k$ is the wavenumber and $d$ is the water depth.
The envelopes, obtained from the Hilbert transform, of the heave and surge displacements of the buoy around the giant wave are shown in  Fig.~\ref{xzratio}a.
The ratio of the surge to heave envelopes, $|s_x/s_z|$, is plotted in Fig.~\ref{xzratio}b alongside $\coth(kd)$, with $k$ calculated from the zero-upcrossing wave periods. 
The ratio of the envelopes has not been smoothed.
The ratio of the envelopes is lower than the linear ratio $\coth(kd)$ at around $t = 0$, suggesting that the local motions are not consistent with 
simple linear wave theory.
{The ratio, for nonlinear waves, would be lower due to the elevated crests, but not as low as we observe for the giant wave. 
For displacement time series recorded by a moored buoy, linear theory should provide a slightly conservative test.}

To better understand the motion of the buoy around the giant event, the instantaneous phase of the surge displacement relative to the heave, obtained via the Hilbert transform, is plotted in {Fig.~\ref{xzratio}c}.
All phase angles have been wrapped around $[-3\pi/2, \pi/2]$.
For a clockwise orbital motion, given a coordinate system with vertical axis positive upwards, and horizontal axis positive in the direction of wave propagation, horizontal motion lags the vertical by $\pi/2$.
However,  around the giant event the relative phase changed from $-\pi/2$ to nearly $-3\pi/4$, 
consistent with the surge-heave trajectory of the buoy shown in Fig.~\ref{orbit}, 
then back to $-\pi/2$ at approximately $50 < t < 100$ s, before
{fluctuating}
for a brief period at $100 < t < 150$ s.


\subsection{Buoy motion trajectories}

\begin{figure*}
\centering
\includegraphics[width=33pc]{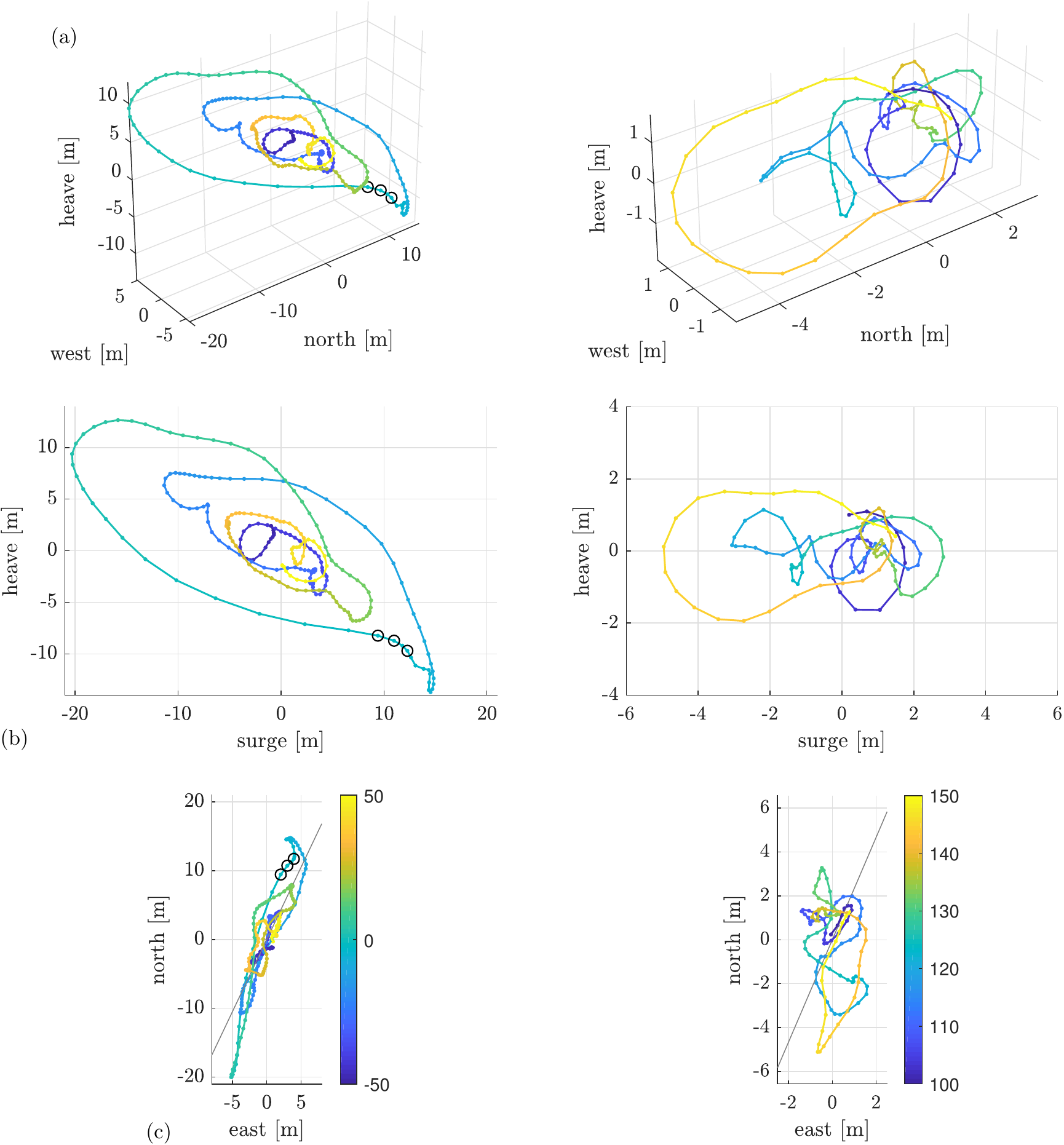}
\caption{Trajectories of buoy displacement for $-50 \leq t \leq 50$ s (left column) and for $100 \leq t \leq 150$ s (right column): (a) perspective view, (b) as seen from the side along the mean wave direction, and (c) from the top. 
Colour indicates time in seconds. 
Time interval between points is $\Delta t = 1/2.56 = 0.39$ s.
Instances of missing data points are indicated with open circles. 
In (c), the mean direction of the waves is indicated by the grey line.
}
\label{orbit}
\end{figure*}

The trajectories of the buoy, as viewed from the top and from the side, along the mean wave direction, are plotted in Fig.~\ref{orbit}. 
The perspective view is also shown.
The trajectories on the left are for a 100-second duration centred at the zero-upcrossing of the largest heave displacement.
The heave-surge trajectory is clearly not in the form of 
an ellipse with a horizontal major axis,
as would be expected from a wave with a sinusoidal profile, but instead is tilted downwards{, consistent with the relative phase observation in Fig.~\ref{xzratio}c}. 
{We have not seen this tilted trajectory reported elsewhere, nor have we found similar trajectories in our dataset.
Studies on particle trajectories under breaking waves~\citep{lenain_pizzo_melville_2019,Chen2012} indicate trajectories  aligned largely along the horizontal.}

The corresponding trajectories for $100 \leq t \leq 150$ s, a few waves after the giant event, is shown on the right side of Fig.~\ref{orbit}. 
At some point, we may notice that the buoy was not making a looping clockwise motion, but was moving up and down while moving backwards. 
{The observed trajectory has the characteristics of a surface particle trajectory at the onset of breaking, for which the upper half of the trajectory is largely elliptical while the lower half has two upward pointing cusps~\citep{Chen2012}. 
The particle would go on to experience a large jump downwave during breaking, with a horizontal displacement on the order of the wavelength~\citep{lenain_pizzo_melville_2019}. 
A moored buoy, however, could not be advected a significant fraction of a wavelength; instead it would return towards its equilibrium position under the restoring force of the mooring line.}

\subsection{Numerical reconstruction}

\begin{figure*}
\centering
\includegraphics[trim = 1mm 3.4mm 4mm 5mm, clip, width=33pc]{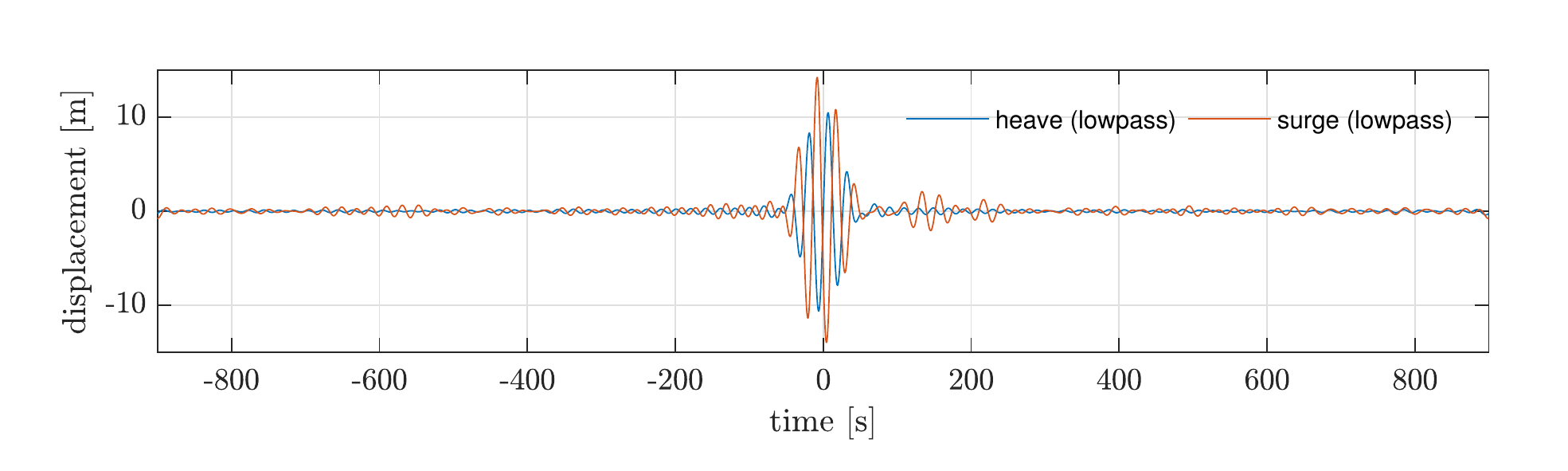}
\caption{Low-pass-filtered surge and heave displacements of the buoy.}
\label{lowpass}
\end{figure*}

\begin{figure*}
\centering
\includegraphics[trim = 1mm 3.4mm 4mm 4mm, clip, width=33pc]{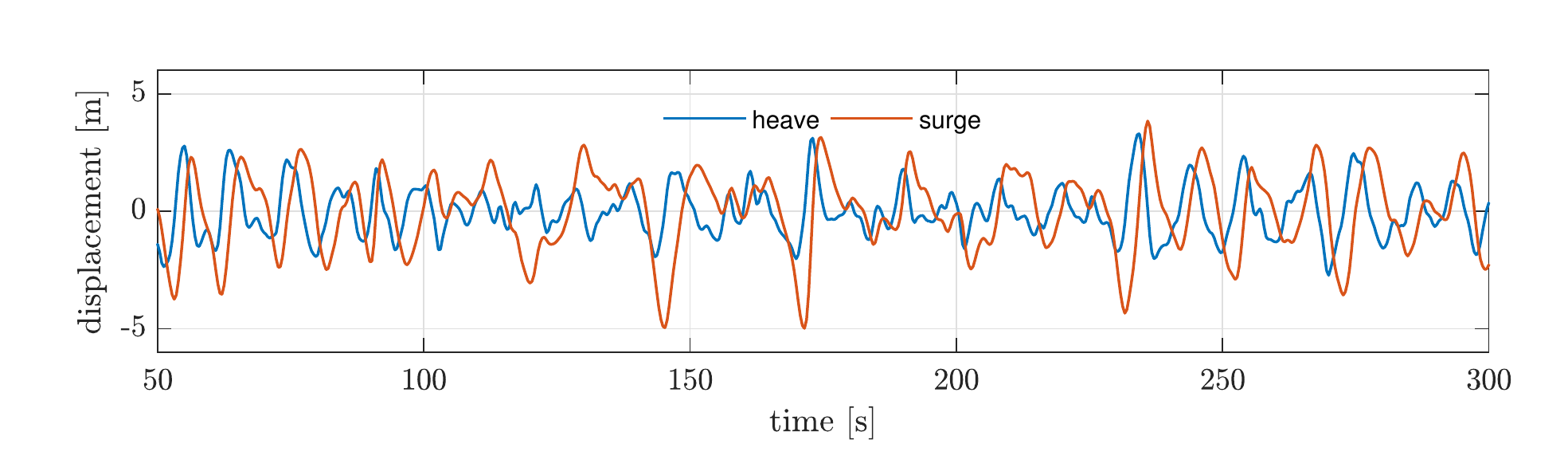}
\caption{Heave and surge displacements of the buoy from $t = 50$ s to $t = 300$ s.}
\label{giant_50s-300s}
\end{figure*}

The time interval $100 < t < 150$ s roughly coincides  with the interval at which a low-frequency surge displacement pulse was recorded some time after the giant wave group (Fig.~\ref{lowpass}).
These low-pass-filtered displacements  have been obtained after removing all frequency components above $1/20$ Hz, i.e.~keeping only long-period oscillations with period greater than 20 s.
A close up view of the raw buoy displacements around this time (Fig.~\ref{giant_50s-300s}) reveals long-period surge oscillations with seaward excursions greater than the landward ones.
This low-frequency pulse is {interesting since
the time at which this pulse was recorded seems to be a plausible range of times for a reflection from the shore to return to the buoy, given the distance of 1 km between the buoy and the shore.}

To investigate how the giant wave group would interact with the coast, and to verify whether the smaller recorded surge pulse could be due to wave reflection from the cliff, a simulation is performed using OXBOU, a numerical model based on the Boussinesq and the nonlinear shallow water equations in 1DH (one horizontal dimension) \citep{ORSZAGHOVA2012328}.  
{Boussinesq models are commonly used in coastal engineering for wave propagation in intermediate and shallow water~\citep[see, e.g.,][]{Brocchini2013} with several variants also suitable for deep water conditions~\citep[see, e.g.,][]{agnon1999,madsen2002}.
The 1DH simulation using a unidirectional wave group, instead of weakly-spread conditions, would result in somewhat exaggerated runup---see~\citet{hunt2003extreme} who reports  approximately 20\% reduction in run-up for a $\pm 10^\circ$ uniformly spread sea---as well as a larger reflection from the coastline, due to lower incident set-down in spread conditions~\citep[see][]{JUDGE2019103531}---and so provide a conservative estimate for the offshore propagating energy content.}

\begin{figure*}
	\centering
\includegraphics[width=39pc]{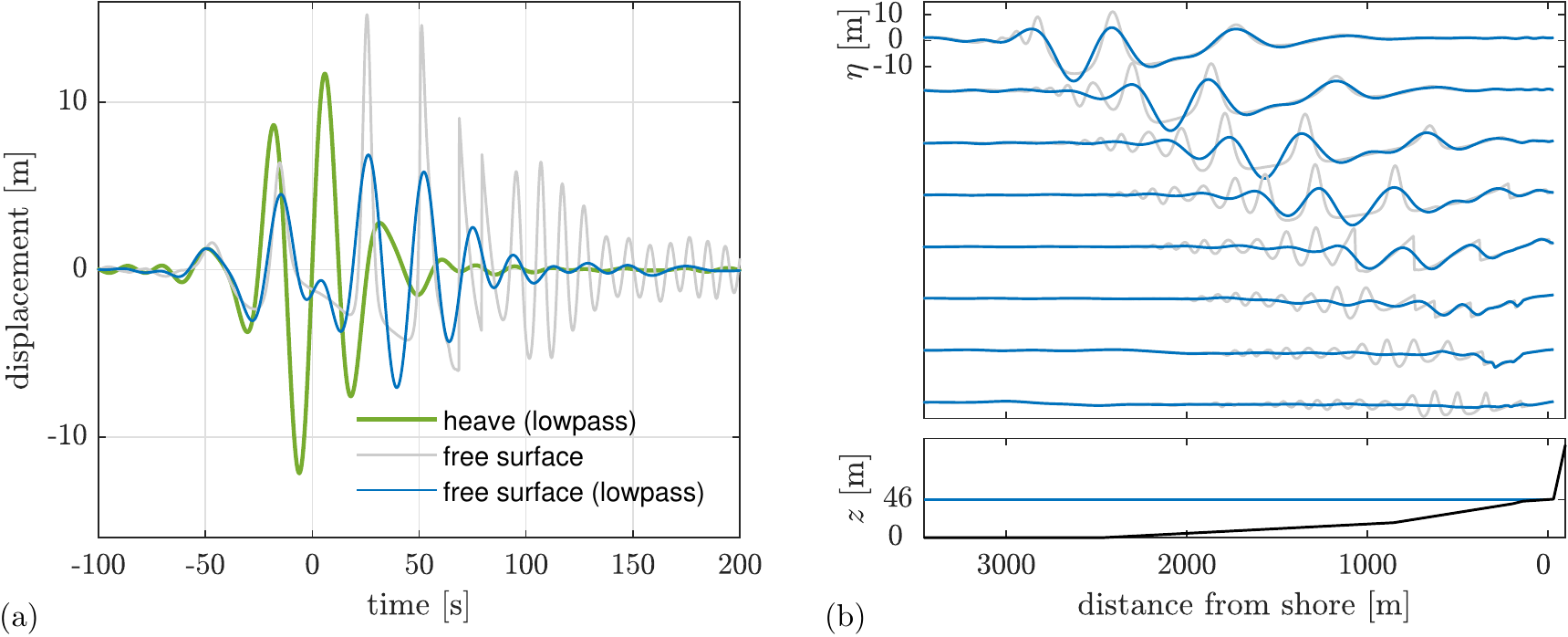}
	\caption{(a) Time series of the measured heave motion and the simulated free surface at the buoy location. (b) Stacked free surface profiles (30 s apart) corresponding to the model bathymetry shown beneath.}    
	\label{Bouss}
\end{figure*}

The bathymetry transect is obtained from a 2018 LiDAR survey of the area. The numerical domain extends to roughly 3.5 km offshore, with offshore water depth of 46 m (Fig.~\ref{Bouss}). The wave group is introduced into the domain via a numerical paddle. The piston paddle motion is calculated from the low-pass filtered heave timeseries: the phase of each frequency component is adjusted by accounting for linear dispersion from the paddle location (3.5 km offshore) to the buoy location (1 km offshore). The actual paddle motion is calculated to second order in order to avoid contamination arising from error waves \citep{ORSZAGHOVA201463}. In essence, the model simulates nonlinear propagation of the wave group into shallower water forced by offshore boundary conditions, which follow from the measured wave field that has been moved offshore linearly. 

The measured heave time series and the simulated free surface at the buoy location are compared in Fig.~\ref{Bouss}. The heave motion is not well reproduced, which is perhaps not surprising given the simple way in which the model forcing is calculated. We note that due to nonlinear wave transformation, as the waves propagate into shallower water, it does not appear possible to recreate a free surface at the buoy location which matches the low-pass filtered heave time series. 
Nonetheless, it is interesting to observe the behaviour of the simulated wave group. The spatial profiles of the simulated free surface are shown in Fig.~\ref{Bouss}, where both the full (grey) and the low-pass filtered (blue) profiles are displayed. The nonlinear wave profiles, resembling cnoidal waves, can be seen. Wave breaking is predicted to occur just past the wave buoy location (1 km offshore). The group is largely dissipated in the surf zone with minimal reflection detected. The vertical runup on the cliff was simulated to be just over 8 m, which is fairly significant. Presumably such a large runup event would result in disturbances to either the vegetation on the cliff face and/or the loose rocks on the beach. However, no discernible differences were observed in images captured on the day of the giant wave recording compared to the days before and after. The images are from an existing onshore camera installed on the cliff overlooking the coast, which records a single image each day at around 15:00 in the afternoon (the event was recorded at around 10:30 in the morning).
{This and the difficulty in reconstructing the giant wave event point to the improbability of the event being real.}




\section{Sea states where accelerometers reached maximum limit}
\label{sec:accelerometers}

\begin{figure*}
\centering
\includegraphics[width=39pc]{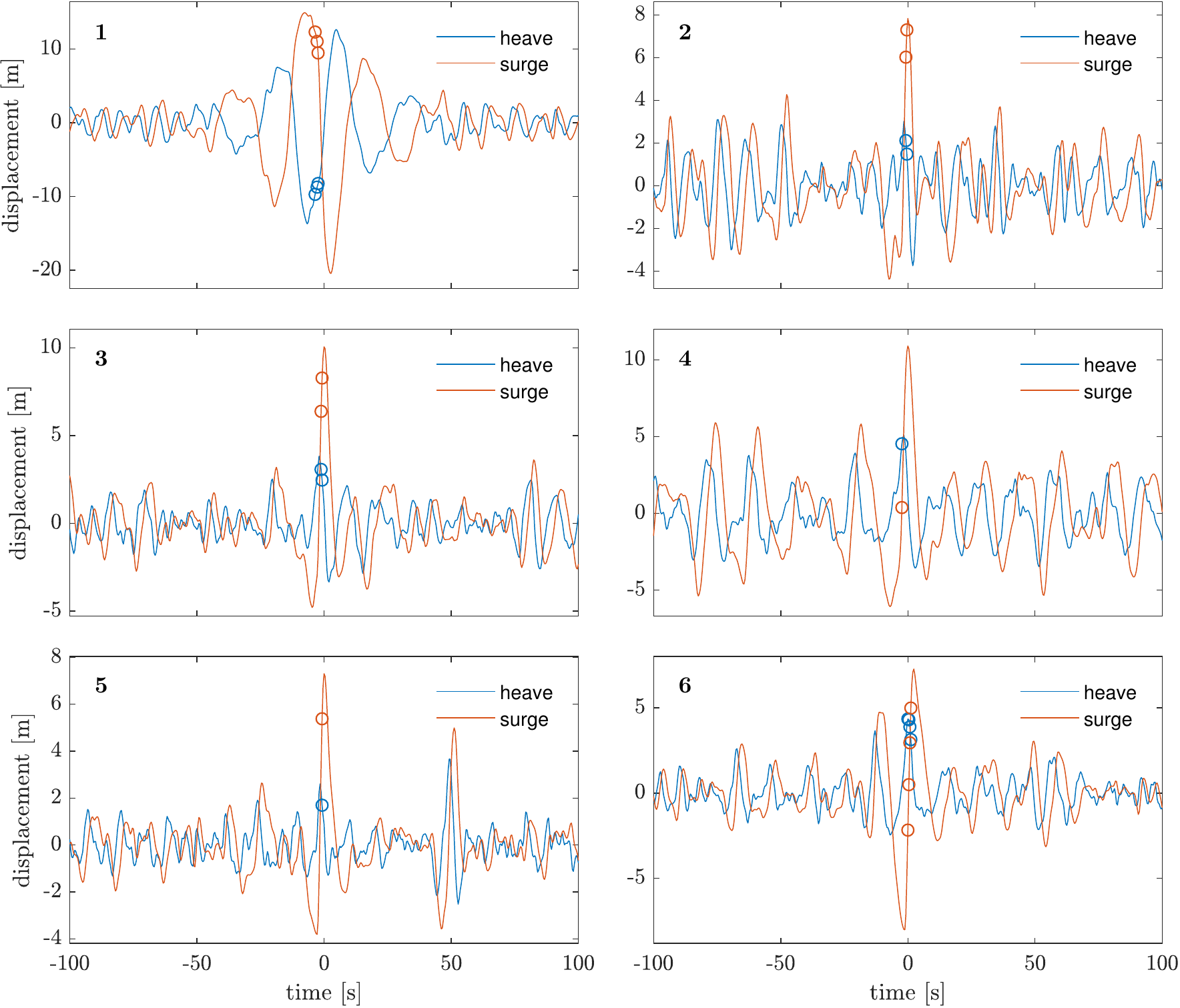}
\caption{Instances of the missing data points indicated on the heave and surge displacements, for the six sea states where at least one accelerometer reached its maximum limit.}
\label{missing}
\end{figure*}

It is mentioned in Section~\ref{secstatistics} that within the one-year record there are six sea states with missing displacement data points.
One of these sea states is the sea state containing the giant wave group, where the instances of the missing data points are shown in the top left plot of Fig.~\ref{missing}.
These are also indicated in the trajectory plots of Fig.~\ref{orbit}.
Examination of the system message file further reveals that 
within the same 30-min sea state,  
the $y$-axis accelerometer reached its maximum limit three times. 

It has been pointed out that spikes in the buoy accelerations~\citep{Cahill2013,soton79448} could manifest as false giant events in the displacement time series, and so we examine the data to identify instances when at least one of the accelerometers reached its limit.
The acceleration time series are not available to us, but the number of times the accelerometers reached their limit is recorded in the system message files{, although the exact time instances when this happened are not known}.
From examination of all the system message files, we discover that within the one-year record
there are in total six sea states in which at least one of the accelerometers reached its limit. 
The 30-min sea states in which these occurred are found to be the same as the aforementioned sea states with missing data points. 
In four of these six sea states, the total number of times the accelerometer reached its limit is equal to the number of missing data points, 
but for two of the sea states (Events 3 and 5 in Table~\ref{events}), there are only two and one missing data points whereas the accelerometer reached its limit four and two times, respectively. 
The instances where the data points are missing in these six sea states are shown in Fig.~\ref{missing}.

\begin{table*}
\caption{Description of sea states where at least one accelerometer reached its maximum limit.}
\label{events}
\begin{center}
\begin{tabular*}{\textwidth}{@{\extracolsep\fill}ccccccccp{46mm}}  
\topline
Event    & Date \& time & $H_\mathrm{mean}$ [m] & $H_s$ [m] & $H_\mathrm{max}$ [m] & $T_z$ [s] & $T_p$ [s] & $T_\mathrm{max}$ [s] & Description \\
\midline
1      & 19 Sep 2019, 10:30    & 3.05 & 6.88 & 21.25 & 8.03 & 26.48 & 26.84 & Giant wave event; $y$-axis accelerometer reached maximum limit 3 times     \\
2      & 19 Sep 2019, 07:00    & 2.66 & 4.40 & 6.77 & 7.74 & 9.89 & 15.23 & $x$-axis accelerometer reached maximum limit 2 times     \\
3      & 04 Sep 2019, 07:30    & 2.57 & 4.42 & 7.33 & 7.66 & 14.20 & 15.62 & $x$-axis accelerometer reached maximum limit 4 times     \\
4      & 22 Jul 2019, 00:00     & 3.52 & 6.37 & 9.38 & 9.56 & 16.45 & 20.10 & $x$-axis accelerometer reached maximum limit 1 time      \\
5      & 05 Nov 2018, 07:00    & 1.80 & 3.30 & 6.59 & 6.54 & 13.13 & 17.31 & $x$-axis accelerometer reached maximum limit 2 times     \\
6      & 22 Oct 2018, 05:30    & 2.30 & 3.97 & 6.13 & 7.96 & 13.59 & 15.08 & $x$-axis accelerometer reached maximum limit 3 times, vertical accelerometer reached maximum limit 1 time, pitch angle reached maximum limit 1 time    \\
\botline
\end{tabular*}
\end{center}
\end{table*}

A common feature in all these six sea states is the large surge displacement of the buoy around the instant of the missing data points. 
However, apart from the giant wave event, the waves recorded around the missing data points (as inferred from the heave displacements of the buoy) in the other five sea states do not look exceptionally different from the background waves. 
It is notable that for the giant wave event (Event 1) the large wave appears more as part of a group rather than an isolated wave as in most of the other events.
We also note that the giant wave event was the only event where the $y$-axis accelerometer reached its maximum limit. 
In the other events, it was either the $x$-axis accelerometer or the vertical axis accelerometer that reached its limit.

The six sea states are listed in Table~\ref{events} together with calculated values of the mean wave height $H_\mathrm{mean}$, significant wave height $H_s$, maximum wave height $H_\mathrm{max}$, mean zero-upcrossing period $T_z$, peak period $T_p$, and maximum period $T_\mathrm{max}$. 
The peak period $T_p$ is obtained from the estimated spectrum of the corresponding heave displacement. 
The $H_\mathrm{max}/H_s$ and $T_\mathrm{max}/T_z$ ratios corresponding to these sea states are marked in Fig.~\ref{HTratio}. 
Apart from the giant wave event, 
the occurrence of the maximum wave height in each of these events is consistent with the expected statistics from a Rayleigh distribution within the sea state. 
Observation of the scatter plot (Fig.~\ref{scatter_plot}) indicates that all these sea states have a low probability of occurrence and relatively steep waves. 

\begin{figure*}
\centering
\includegraphics[width=39pc]{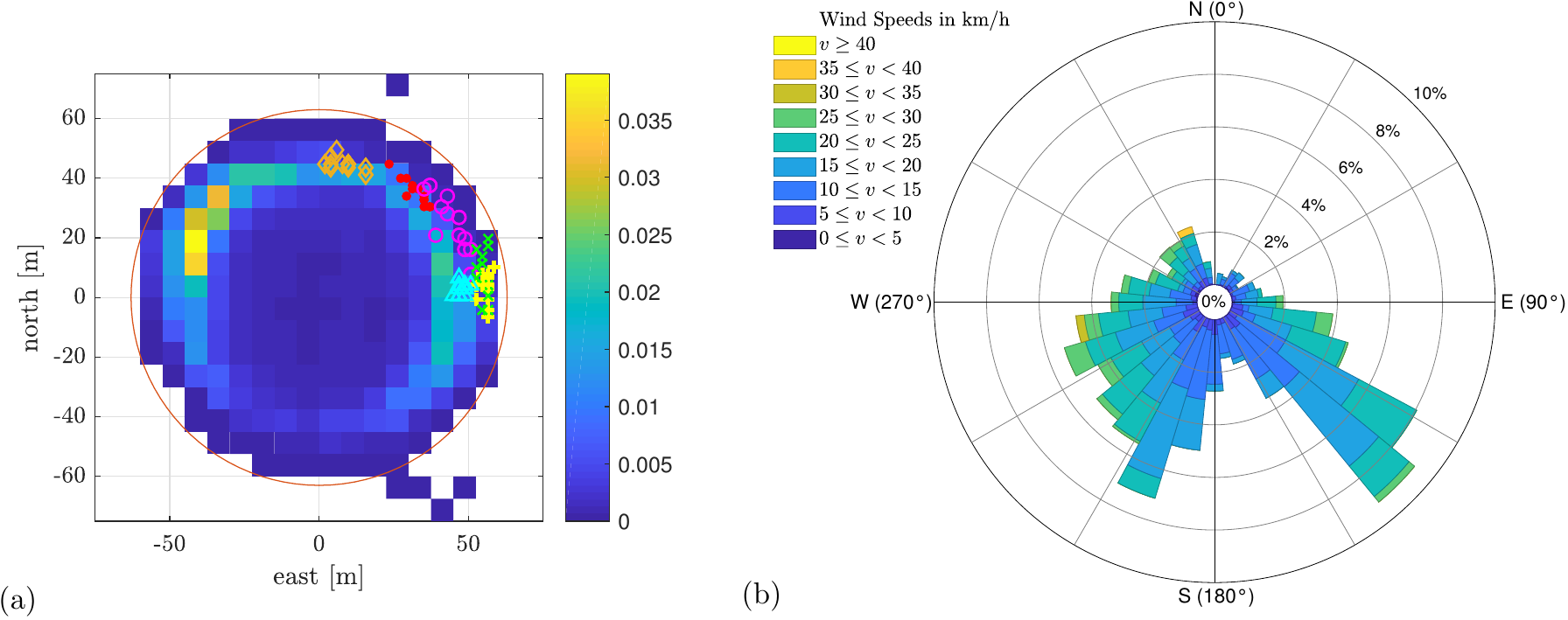}
\caption{(a) Histogram of the buoy GPS location within the year. A watch circle of radius 63 m is also shown. 
The estimated locations of the buoy for 1-hour duration around the six events of Table~\ref{events} are indicated by the symbols: 1 ($\cdot$), 2 ($\circ$), 3 ($+$), 4 ($\times$), 5 ($\triangle$), 6 ($\diamond$). (b) Wind rose for the 3 pm hourly wind speed and direction records measured at Albany airport~\citep{albanyairport}, approximately 14 km inland from the buoy location, for the one-year period corresponding to the buoy record.}
\label{gpswindrose}
\end{figure*}
\begin{figure}
\centering
\includegraphics[width=15pc]{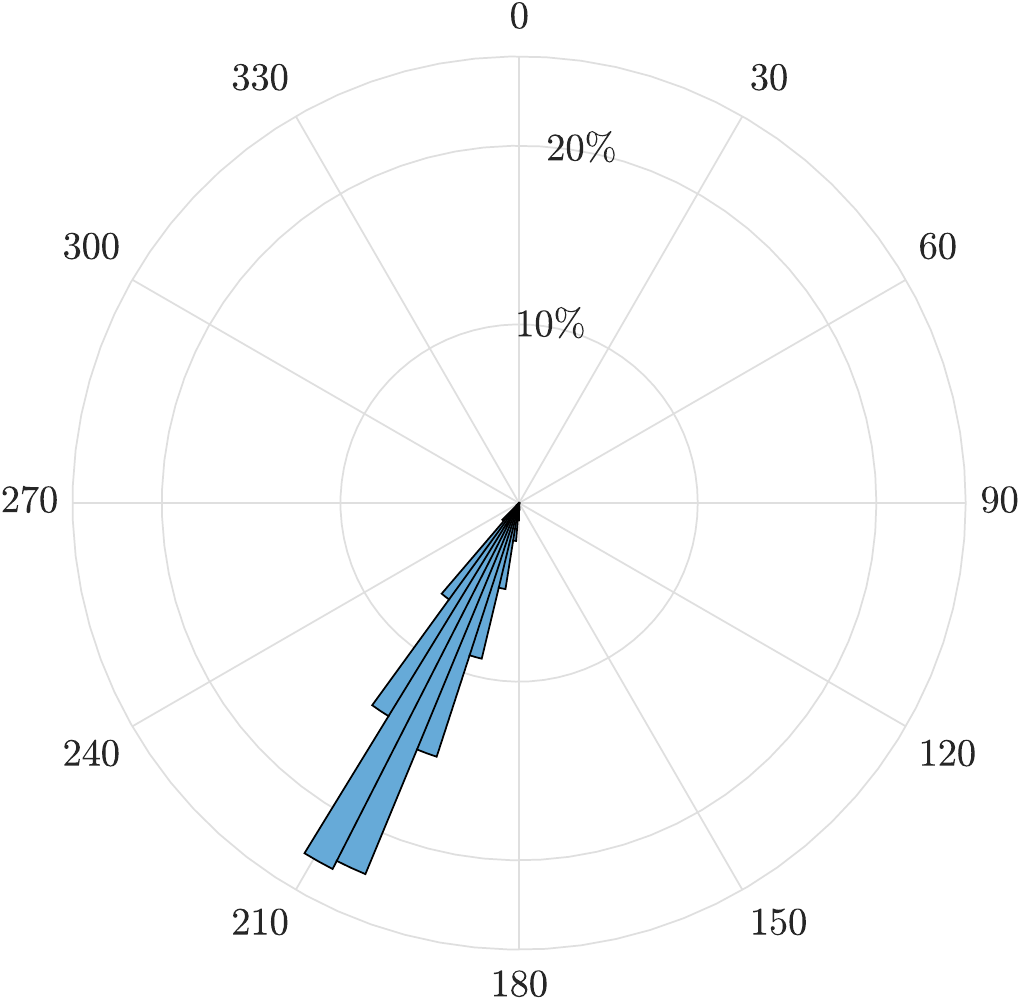}
\caption{Polar histogram of the mean wave direction for the one-year period.}
\label{dirrose}
\end{figure}

A histogram of the GPS location of the buoy for the entire year is shown in Fig.~\ref{gpswindrose}a.
The estimated locations of the buoy during the six sea states are indicated by symbols.
A watch circle of radius 63 m, calculated based on
the water depth of 30 m and an assumed mooring line length  of 70 m (neglecting any stretch), is also shown.
From the histogram, it is seen
that the buoy spent most of its time at the west north west direction relative to its geometric mean. 
This {correlates} with the frequent afternoon sea breezes from the south-east (see Fig.~\ref{gpswindrose}b). 
For all the six sea states with missing data points, the buoy was in the first quadrant (between north and east direction) and relatively close to the watch circle. 
The predominant wave direction throughout the year was from approximately 210$^\circ$ measured from north (see Fig.~\ref{dirrose}).

\begin{figure*}
\centering
\includegraphics[width=39pc]{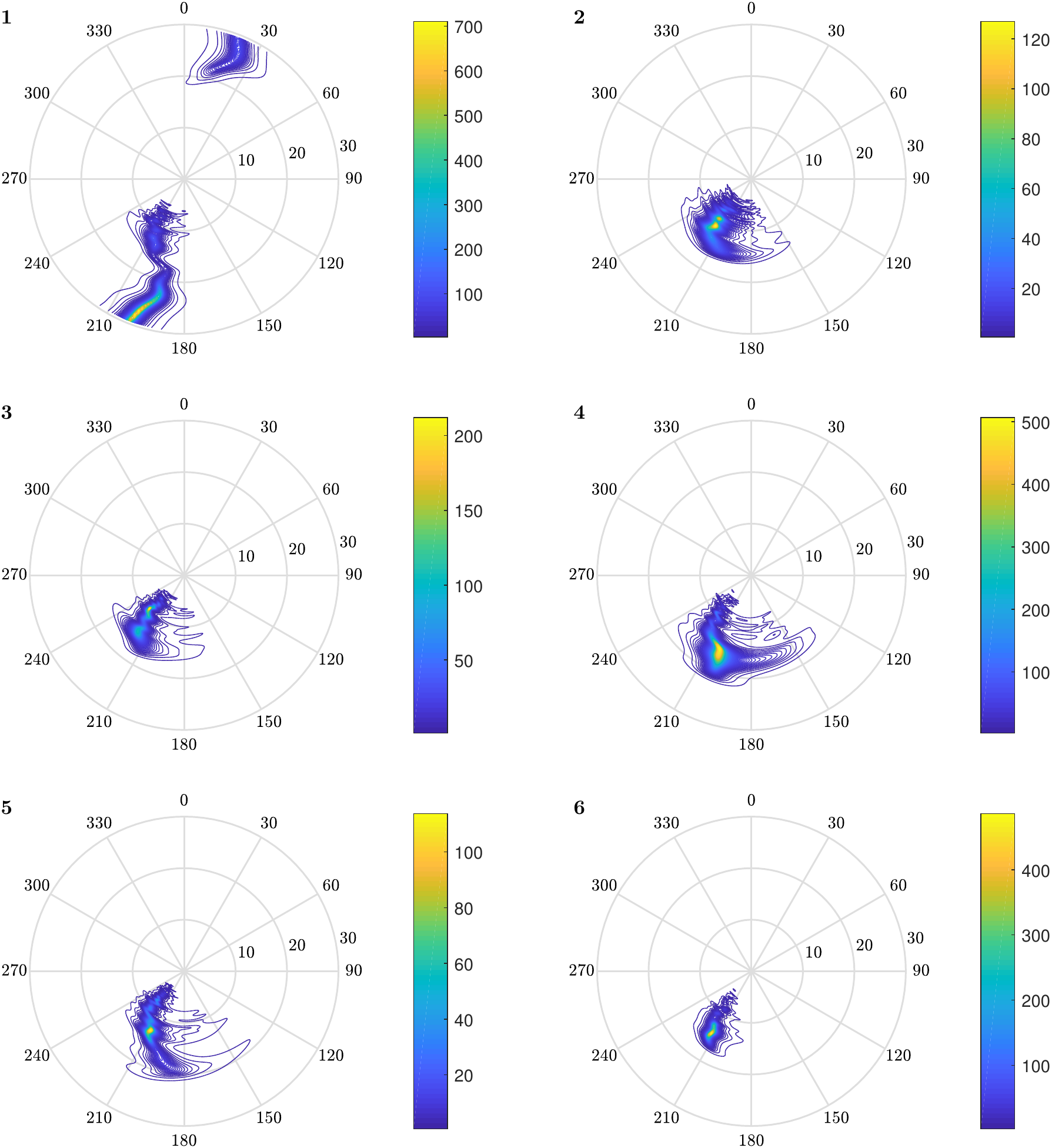}
\caption{Directional spectra (in m$^2$ s rad$^{-1}$) of the sea states where at least one accelerometer reached its maximum limit. Radial coordinates denote wave periods in seconds. Angular coordinates denote the direction (in degrees) from which the wave propagates. For each plot, 200 contour levels are used.}
\label{Sdir}
\end{figure*}

The directional spectra of the six sea states are shown in Fig.~\ref{Sdir}.
They have been estimated using the maximum entropy method following~\citet{LygreKrogstad1986}.
Rather than using 30-min long records, 1-hour long records of the heave, north, and west displacements, centred at the event of interest, are used in this estimation.

An interesting detail emerges from this analysis.
The directional spectrum of the sea state containing the giant wave shows some energy content in the direction opposite the main wave propagation direction. 
This is not present in any of the other five sea states nor in any of the other sea states in the entire year. 

\begin{figure}
\centering
\includegraphics[width=19pc]{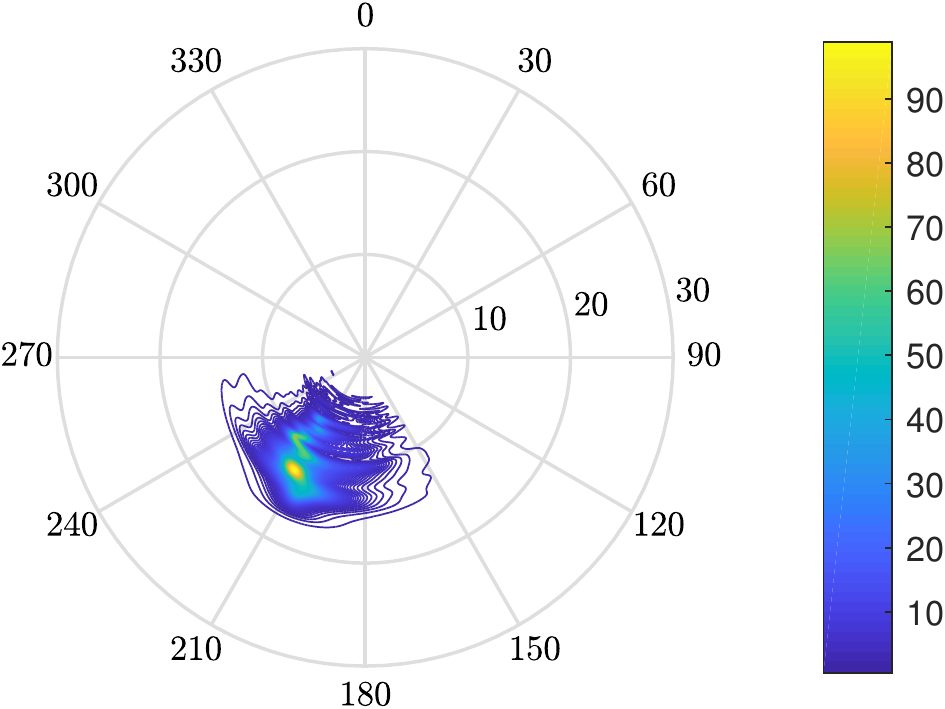}
\caption{Directional spectrum (in m$^2$ s rad$^{-1}$) of {the sea state containing the giant wave (sea state 1 in Fig.~\ref{Sdir})} with the giant wave group ($-50 < t < 50$ s) removed. 
}
\label{Sdir1squashed}
\end{figure}

To investigate this more closely, we ramp down the heave, north, and west displacement time series to zero for $-50 < t < 50$ s, thus removing the giant wave group, and then recompute the directional spectrum. 
The resulting directional spectrum with the giant wave group removed is shown in Fig.~\ref{Sdir1squashed}.
The spectral components with $T > 20$ s disappear, including the components propagating in the opposite direction, demonstrating that they arise from the giant wave group itself and not from the long-period surge displacement pulse recorded at approximately $100 < t < 150$ s and suspected to be reflection from the beach, discussed earlier.

We also perform the same procedure on the other five sea states, where the displacements at $-20 < t < 20$ s around the centre of the events are ramped down to zero. 
The resulting spectra are not plotted here, but they {are} similar to the original spectra shown in Fig.~\ref{Sdir}.  

\section{Additional analysis} \label{secaddanalysis}

The Western Australian Department of Transport operates a wave buoy at 35$^\circ$11'53.0''S 117$^\circ$43'19.0''E, roughly 15 km upwave of our buoy along the predominant wave direction.
This is a Datawell Waverider MkIII buoy \citep{DWRMkIIImanual}, which is also an accelerometer-based wave buoy. 
A comparison with data from this buoy would have been helpful to validate the giant event, but this
was not possible as there was no record available on the date {of the event},
possibly because it was overwritten. 
Since the overwriting rule is based on the value of the significant wave height, this suggests that the underlying sea state where the giant event might be recorded was otherwise relatively moderate. 

\begin{figure*}
\centering
\includegraphics[width=39pc]{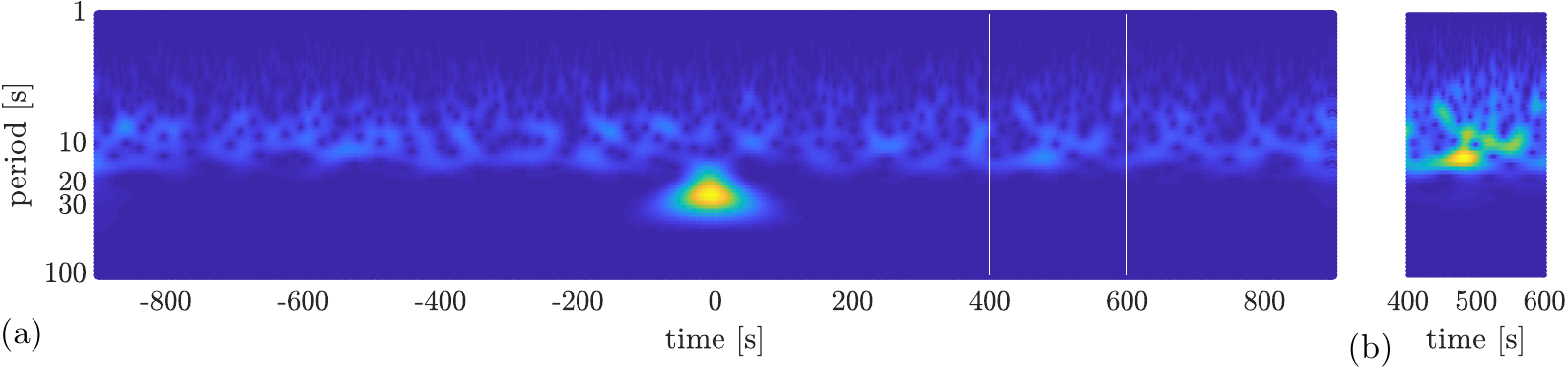}
\caption{Wavelet transform of the heave displacements of the buoy for (a) $-900 \leq t \leq 900$ s and  (b) $400 \leq t \leq 600$ s {(plotted on a different colour scale).}}
\label{waveleth}
\end{figure*}
\begin{figure*}
\centering
\includegraphics[width=33pc]{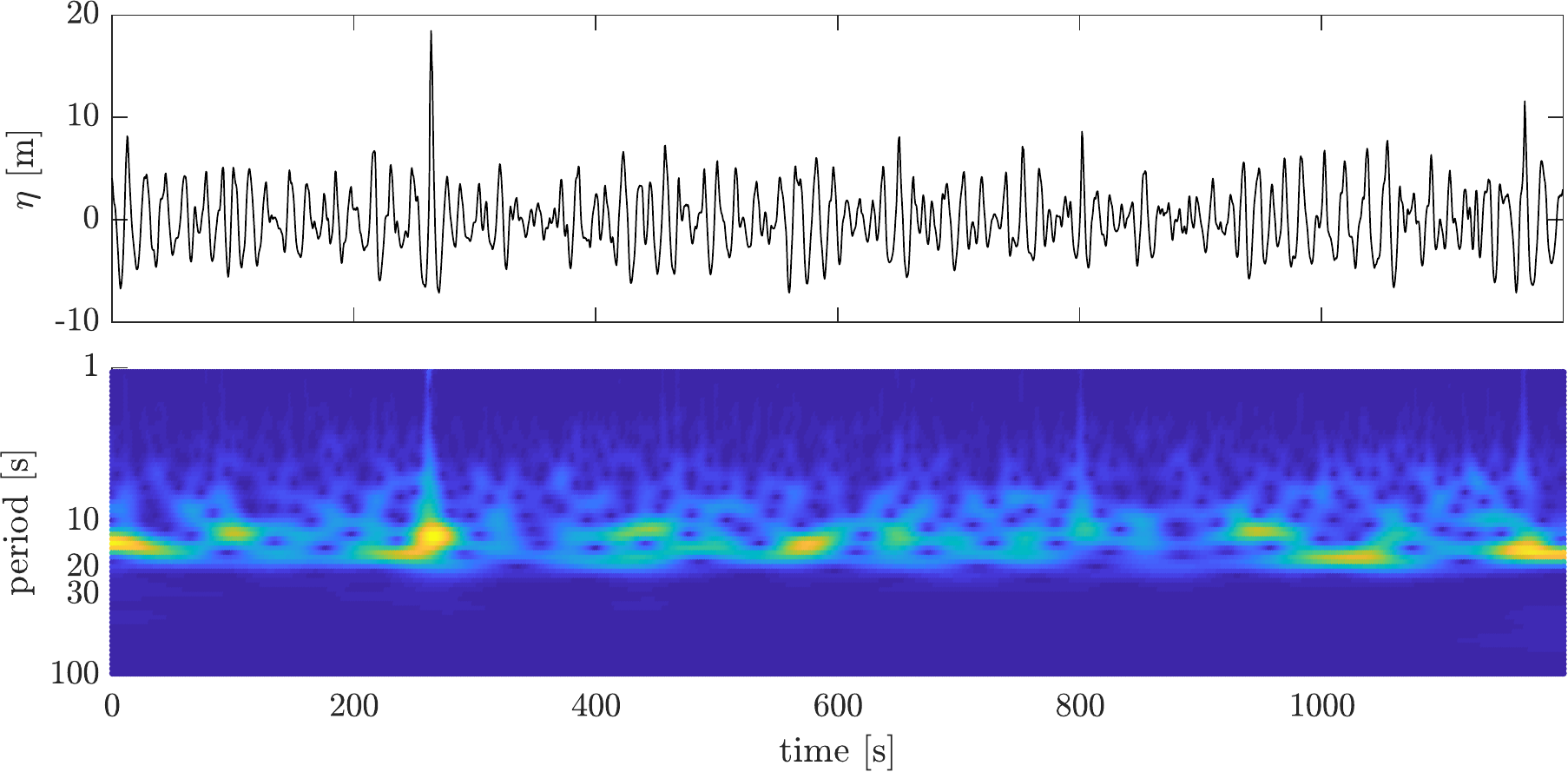}
\caption{{Free-surface elevation record (20 minutes) containing the Draupner wave, and its wavelet transform.}}
\label{waveletdraupner}
\end{figure*}

As a further investigation, wavelet analysis is performed on the displacement time series. 
{The use of wavelet transform as a tool to detect anomalies has been suggested, e.g., by~\citet{Cahill2013,MORI20021399}.}
Wavelet transform of the heave displacements of the buoy for the 30-minute record centred at  the giant event, 
obtained using the analytic Morse wavelet~\citep{1041025}, is shown in
Fig.~\ref{waveleth}a.
The giant wave shows up {prominently} in the wavelet spectrum,
{with low frequency content not present elsewhere in the 30-minute record.
For comparison, the wavelet transform for the interval 400--600 s, which contains the second highest wave outside the giant wave group within the same  record, is shown in Fig.~\ref{waveleth}b.}
{We can also contrast this to the wavelet spectrum of the Draupner wave (Fig.~\ref{waveletdraupner}), which is similarly characterised by a strong energy density at the onset of the wave, but with dominant frequency content aligned with that of the rest of the sea state.}

\begin{figure*}
\centering
\includegraphics[width=39pc]{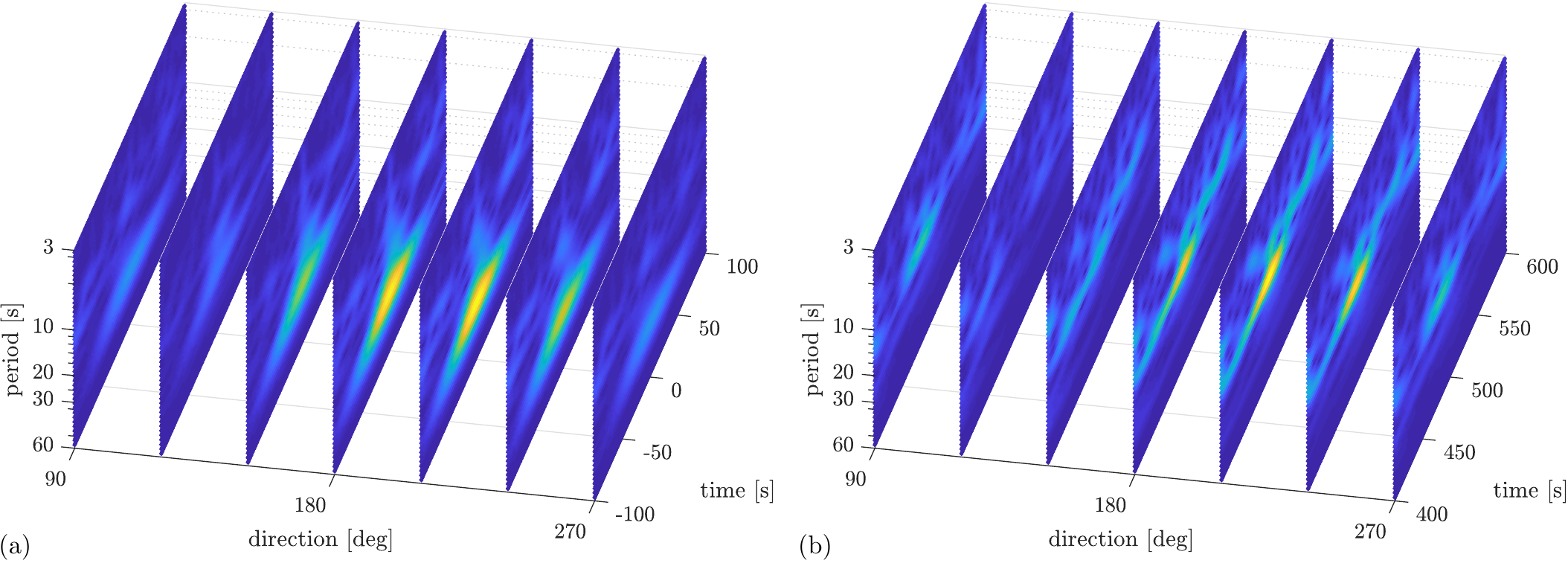}
\caption{Wavelet transforms of the horizontal displacements of the buoy, resolved along different directions, at (a) around the giant wave event ($-100 \leq t \leq 100$ s) and at (b) $400 \leq t \leq 600$ s.}
\label{wavelets}
\end{figure*}

Fig.~\ref{wavelets} shows the wavelet transforms of the horizontal displacements of the buoy along different directions, at around the giant wave and at $400 \leq t \leq 600$ s, for comparison. 
The giant event (as long-period oscillations) is evident in all directions, but is most pronounced at around the mean direction of the waves, 210 degrees.

\begin{figure*}
\centering
\includegraphics[width=33pc]{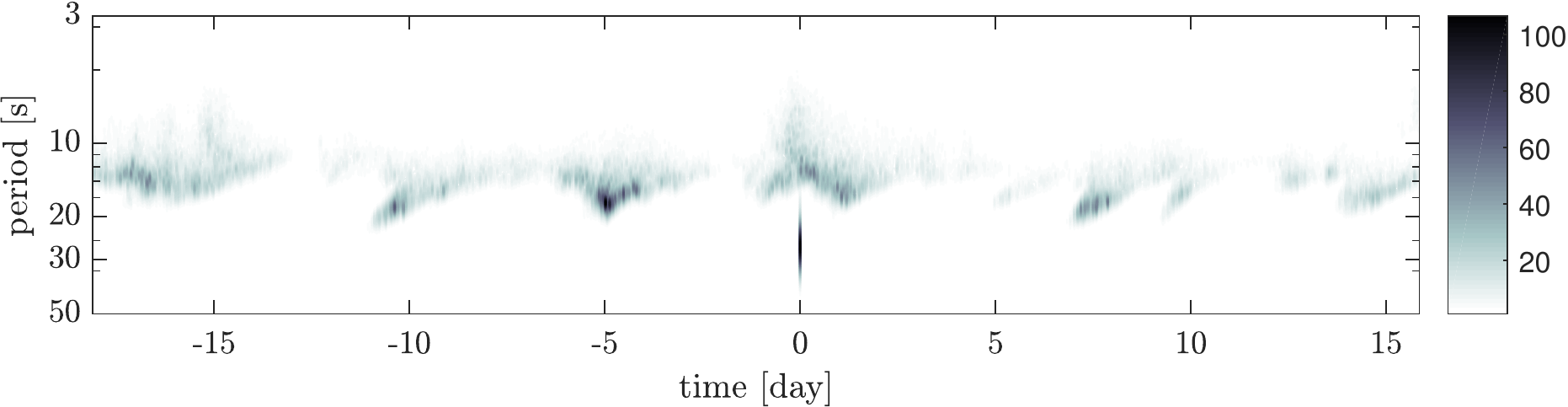}
\caption{Transformation of the heave displacement spectrum (in m$^2$ s) over time, centred at the giant event.}
\label{Shhtransformation}
\end{figure*}

A similar investigation is performed over a longer time scale by plotting the heave displacement spectrum as a function of time (Fig.~\ref{Shhtransformation}). 
Arrivals of swell from  distant storms, starting with longer period waves which propagate faster and thus arrive earlier, followed by shorter period waves which propagate slower and arrive later, are evident. 
The distance to the storm centre can be estimated by observing the slope of each streak~\citep{doi:10.1098/rsta.1948.0005}. 
Assuming deep water dispersion, this ranges from about 3500 km to 7500 km.
The sea state containing the giant event shows up quite remarkably in this plot as well and looks markedly different from the rest of the sea states.
The mechanism behind its generation does not seem to be normal.

\section{{Recommended detection tests}} \label{sectests}

\begin{figure}
\centering
\includegraphics[width=27pc]{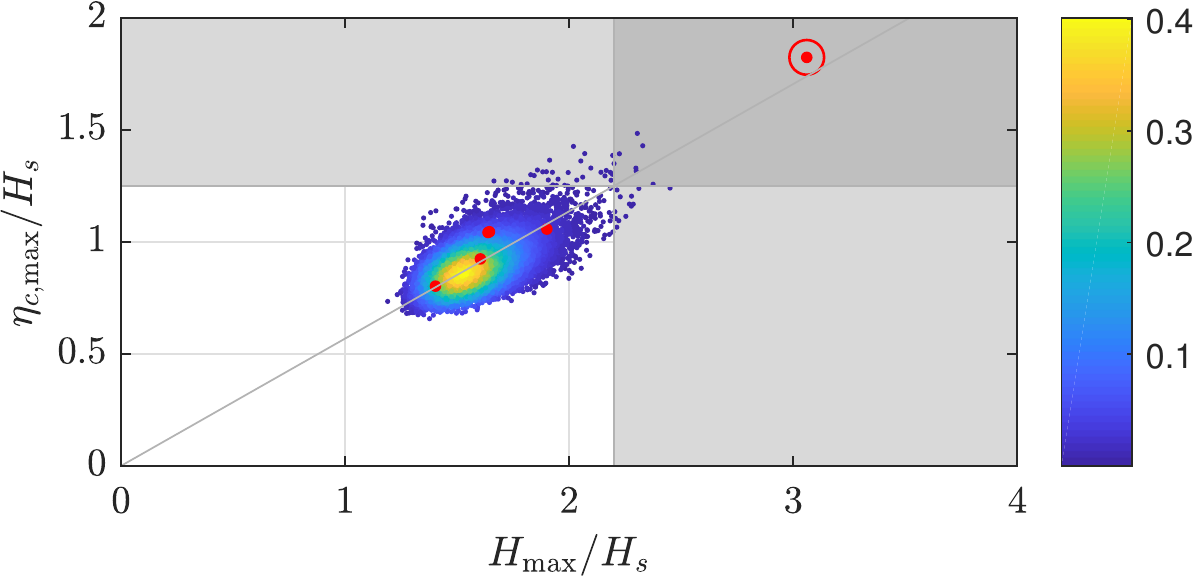}
\caption{{All 30-minute sea states in the one-year record plotted in the $(H_\mathrm{max}/H_s, \eta_{c,\mathrm{max}}/H_s)$ space. 
Color indicates point density.
The sea states in which at least one of the accelerometers reached its maximum limit are marked in red.
The sea state in which the giant wave event occurred is enclosed with a red circle. 
Shaded areas indicate the commonly adopted rogue wave criteria, $H_\mathrm{max} > 2.2 H_s$ or $\eta_{c,\mathrm{max}} > 1.25 H_s$.}  
}
\label{roguespace}
\end{figure}

{Based on the various analyses described in this paper, a summary of recommended thresholds to detect a false rogue wave event is presented in Table~\ref{tests} and applied to our apparent giant wave event and to the Draupner wave, a real rogue wave event.
In addition, we apply these tests to a number of rogue events in our dataset, which are selected based on the commonly adopted rogue wave criteria of $H_\mathrm{max} > 2.2 H_s$ or $\eta_{c,\mathrm{max}} > 1.25 H_s$ (Fig.~\ref{roguespace}).
For this exercise, we have selected rogue events satisfying both criteria.
Neither Draupner nor the other rogue events from our dataset share any of the extreme characteristics of the giant wave.}

{As a future work, it would be interesting to compare these tests to those suggested in~\citet{BaschekImai2011,Voermans2021}.}

\begin{sidewaystable}
\caption{Summary of detection tests and suggested thresholds. Values for the giant wave, Draupner, and seven rogue sea states from the dataset (RS1--RS7) are in boldface if they meet the thresholds, in regular font if they do not, and italics if no information is available.}
\label{tests}
\begin{center}
\begin{tabular*}{\textwidth}{@{\extracolsep\fill}p{20mm}p{26mm}p{21mm}p{20mm}p{13mm}p{13mm}p{13mm}p{13mm}p{13mm}p{13mm}p{13mm}}  
\topline
Test    & Suggested threshold & Giant wave & Draupner & RS1 & RS2 & RS3 & RS4 & RS5 & RS6 & RS7 \\
\midline
$H_\mathrm{max}/H_s$ & 
$ > 3$ & \textbf{3.06} & 2.15 & 2.20 & 2.30 & 2.38 & 2.33 & 2.30 & 2.31 & 2.23 \\
$\eta_{c,\mathrm{max}}/H_s$ & 
$ > 1.7$ & \textbf{1.82} &  1.55 & 1.35 & 1.32 & 1.26 & 1.43 & 1.33 & 1.48 & 1.39 \\
$T_\mathrm{max}/T_z$ & 
$ > 3$ 
& \textbf{3.31} & 1.42 & 2.05 & 2.11 & 2.33 & 2.60 & 2.12 & 2.37 & 1.87 \\
Least-squares $c$ & $ < 0.001$ & \textbf{0.00081}
& $0.017$ & $0.014$ & $0.014$ & $0.016$ & $0.011$ & $0.013$ & $0.016$ & $0.014$ \\
Expected shape & dissimilarity to the expected shape from neighbouring intervals & \textbf{different} &  similar & similar & similar & similar & similar & similar & similar & similar \\
Surge to heave amplitude ratio & $\ll \coth(kd)$ for a sustained period of more than 50 s  & \textbf{yes} & \emph{not known} & no & no & no & no & no & no & no   \\
Surge to heave relative phase & $\not\approx -\pi/2$ & $\approx \bm{-3\pi}/\bm{4}$ & \emph{not known} & $\approx -\pi/2$ & $\approx -\pi/2$ & $\approx -\pi/2$ & $\approx -\pi/2$ & $\approx -\pi/2$ & $\approx -\pi/2$ & $\approx -\pi/2$ \\
Reconstruction & not possible & \textbf{not possible using Boussinesq model} & possible numerically and experimentally & \emph{not attempted} & \emph{not attempted} & \emph{not attempted} & \emph{not attempted} & \emph{not attempted} & \emph{not attempted} & \emph{not attempted} \\
Accelerometer  & maxing out & \textbf{yes} & \emph{NA} & no & no & no & no & no & no & no \\
Directional spectrum & energy content in opposite direction & \textbf{yes} & \emph{not known} & no & no & no & no & no & no & no \\
Wavelet   & dominant frequency lower than the rest of the sea state  & \textbf{yes} & no & no & no & no & no & no & no & no \\
Spectral transformation & sudden appearance & \textbf{yes} & no sudden signature  & no sudden signature  & no sudden signature  & no sudden signature  & no sudden signature  & no sudden signature  & no sudden signature & no sudden signature  \\
\botline
\end{tabular*}
\end{center}
\end{sidewaystable}

\section{Conclusion}

Oceanographers and offshore engineers are greatly interested in the largest waves, both in a single sea state and also over long periods of time. 
{This has led to much investigation into rogue wave events~\citep[see, e.g.,][and references therein]{Dysthe2008}.
In this paper we have studied a year-long record from a wave buoy off the south-western coast of Australia and investigated in detail an apparent rogue wave event recorded by the buoy.}
{While each of the analysis described herein is suggestive on its own, considering them together we believe rules out the plausibility of the event.} 



\begin{figure}
\centering
\includegraphics[width=27pc]{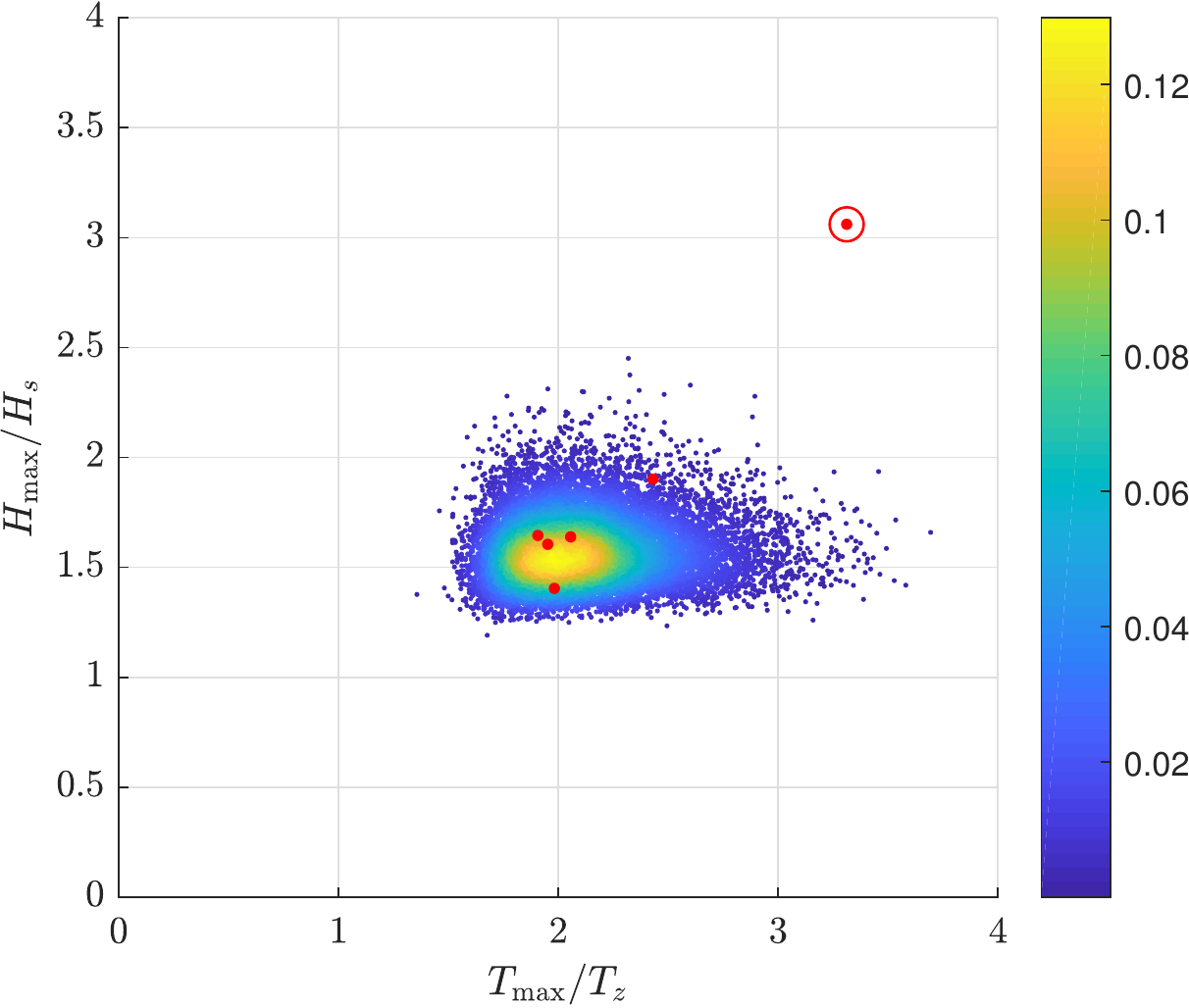}
\caption{{All 30-minute sea states in the one-year record plotted in the $(T_\mathrm{max}/T_z, H_\mathrm{max}/H_s)$ space. 
Color indicates point density.
The sea states in which at least one of the accelerometers reached its maximum limit are marked in red.
The sea state in which the giant wave event occurred is enclosed with a red circle.}  
}
\label{TmaxHmaxspace}
\end{figure}

{The event is an outlier in terms of wave height, with $H_\mathrm{max}/H_s$ ratio as large as 3.06 (in contrast to the average $H_\mathrm{max}/H_s$ of 1.58 calculated from all 30-min sea states in the entire year). 
The probability for the largest wave in any 30-min sea state to exceed this value is in the order of $10^{-6}$ according to the Rayleigh distribution, with which the rest of the sea states appears to be consistent.  
In terms of wave period, while the $T_\mathrm{max}/T_z$ ratio is also high, it does not seem to be a clear outlier when we consider all the sea states in the year.
{However, in the $(T_\mathrm{max}/T_z, H_\mathrm{max}/H_s)$ space, the event is a clear outlier (Fig.~\ref{TmaxHmaxspace}).}
Extending the analysis to the crests and troughs, we found that a simplified second-order model captures the crest-trough asymmetry reasonably well.
{Plotting the best fit second-order coefficient against $k(T_e) d$, where $T_e$ is the energy period,
we found the event to be an outlier.
}
}    

Closer examination of the buoy displacements around the giant wave reveals characteristics which 
{are inconsistent with known behaviour of real waves.}
The ratio of the horizontal to vertical displacements of the buoy around the giant wave is much smaller than expected from theory.
Further, the phase of the horizontal displacement relative to the vertical is shifted by nearly 90 degrees, and the orbital trajectory of the buoy is tilted downward.
{Numerical reconstruction using Boussinesq and nonlinear shallow water wave equations, while not a comprehensive attempt at reproducing the event, could not reproduce the observed characteristics of the giant wave.}

False giant events have been associated with spikes in the acceleration time series, and therefore we have also examined {five other} sea states in the one-year long record with at least one of the three accelerometers reaching its maximum limit.  
We find that all these sea states contain missing data points at around a big event (large horizontal displacement), but apart from the sea state with the giant wave, the vertical displacements in the other sea states are characteristically similar to the background waves.  
It {is clear} that accelerometers reaching their maximum limit does not necessarily manifest as a giant vertical displacement, 
{but it is also clear that such flagged events should be examined closely}.  

{The directional spectrum of the sea state containing the giant wave reveals energy content in the direction opposite the main wave direction.
This has been shown to arise from the giant wave itself, not from an apparent reflection from the coast. 
Further analyses using wavelets and by plotting the displacement spectrum over time  suggest that the giant event did not arise from natural causes.}
%
%
%
%
In contrast, the time signal of the famous Draupner wave 
is consistent with detailed properties of waves in crossing seas~\citep{doi:10.1098/rspa.2011.0049} 
and can be reproduced well in a wave basin~\citep{mcallister_draycott_adcock_taylor_van_den_bremer_2019}.

{It is likely that the recorded event is  an artefact of the digital filter. 
This could be caused by external factors like a collision with a ship or boat, or a breaking wave.
Although we do not completely rule out the former, 
looking at the background sea state in which the event was recorded, $H_s \approx 4.8$ m and $T_z \approx 8$ s (c.f.~Fig.~\ref{scatter_plot}), it is more likely that a breaking wave was the cause of the artefact.}

Historically, wave buoys have been used to estimate wave spectra, both omni-directional and directionally resolved. 
With time series of buoy motion in all three directions now available, potentially much more information should be accessible. 
However, quality assurance of the three simultaneous time series is much more difficult than for standard spectra of vertical motion estimated over many wave periods.  
We recommend that all measurements of rogue or freak waves be subjected to forensic analysis before they are accepted as being real occurrences.

\acknowledgments
This research is carried out as part of {Marine Energy Research Australia}, jointly funded by The University of Western Australia and the Western Australian Government, via the Department of Primary Industries and Regional Development (DPIRD).
Funding was also provided by the Integrated Marine Observing System through their New Technology Proving Program.
The authors acknowledge the Australian Hydrographic Office for the bathymetric LiDAR data, {Thomas Adcock (Oxford) for sharing the Draupner wave time series, Carly Portch (UWA) for providing additional bathymetric data and onshore camera images, and Thobani Hlophe (UWA) for useful discussions concerning Eqs.~\eqref{eqR} and~\eqref{eqL}.
The authors would also like to thank the anonymous reviewers for their valuable comments and suggestions.} 
 
\datastatement
The data that support the findings of this study are available from the corresponding author upon reasonable request.

\bibliographystyle{ametsocV6}
\bibliography{wave}

\begin{thebibliography}{50}
\providecommand{\natexlab}[1]{#1}
\providecommand{\url}[1]{\texttt{#1}}
\renewcommand{\UrlFont}{\rmfamily}
\providecommand{\urlprefix}{URL }
\expandafter\ifx\csname urlstyle\endcsname\relax
  \providecommand{\doi}[1]{https://doi.org/\discretionary{}{}{}#1}\else
  \providecommand{\doi}{https://doi.org/\discretionary{}{}{}\begingroup
  \urlstyle{rm}\Url}\fi
\providecommand{\eprint}[2][]{\url{#2}}

\bibitem[{Adcock et~al.(2011)Adcock, Taylor, Yan, Ma,, and
  Janssen}]{doi:10.1098/rspa.2011.0049}
Adcock, T. A.~A., P.~H. Taylor, S.~Yan, Q.~W. Ma, and P.~A. E.~M. Janssen,
  2011: Did the {Draupner} wave occur in a crossing sea? \textit{Proceedings of
  the Royal Society A: Mathematical, Physical and Engineering Sciences},
  \textbf{467~(2134)}, 3004--3021, \doi{10.1098/rspa.2011.0049}.

\bibitem[{Agnon et~al.(1999)Agnon, Madsen,, and Sch{\"a}ffer}]{agnon1999}
Agnon, Y., P.~A. Madsen, and H.~A. Sch{\"a}ffer, 1999: A new approach to
  high-order {B}oussinesq models. \textit{Journal of Fluid Mechanics},
  \textbf{399}, 319--333, \doi{10.1017/S0022112099006394}.

\bibitem[{Allsop et~al.(1998)Allsop, Durand,, and Hurdle}]{Allsop1998}
Allsop, N., N.~Durand, and D.~Hurdle, 1998: Influence of steep seabed slopes on
  breaking waves for structure design. \textit{26th International Conference on
  Coastal Engineering}, Copenhagen, 906--919, \doi{10.1061/9780784404119.067}.

\bibitem[{Barber et~al.(1948)Barber, Ursell,, and
  Deacon}]{doi:10.1098/rsta.1948.0005}
Barber, N.~F., F.~Ursell, and G.~E.~R. Deacon, 1948: The generation and
  propagation of ocean waves and swell. {I. W}ave periods and velocities.
  \textit{Philosophical Transactions of the Royal Society of London. Series A,
  Mathematical and Physical Sciences}, \textbf{240~(824)}, 527--560,
  \doi{10.1098/rsta.1948.0005}.

\bibitem[{Baschek and Imai(2011)Baschek, and Imai}]{BaschekImai2011}
Baschek, B., and J.~Imai, 2011: Rogue wave observations off the {US} {West
  Coast}. \textit{Oceanography}, \doi{10.5670/oceanog.2011.35}.

\bibitem[{Brocchini(2013)}]{Brocchini2013}
Brocchini, M., 2013: A reasoned overview on {B}oussinesq-type models: {T}he
  interplay between physics, mathematics and numerics. \textit{Proceedings of
  the Royal Society A: Mathematical, Physical and Engineering Sciences},
  \textbf{469~(2160)}, 20130\,496, \doi{10.1098/rspa.2013.0496}.

\bibitem[{{Bureau of Meteorology}(2020)}]{albanyairport}
{Bureau of Meteorology}, 2020: Latest weather observations for {Albany
  Airport}. Accessed: 11-Aug-2020,
  \url{http://www.bom.gov.au/products/IDW60801/IDW60801.94802.shtml}.

\bibitem[{Cahill(2013)}]{Cahill2013}
Cahill, B., 2013: Characteristics of the wave energy resource at the {Atlantic}
  marine energy test site. Ph.D. thesis, University College Cork,
  \urlprefix\url{https://cora.ucc.ie/handle/10468/1142}.

\bibitem[{Chen et~al.(2012)Chen, Li, Hsu,, and Ng}]{Chen2012}
Chen, Y.-Y., M.-S. Li, H.-C. Hsu, and C.-O. Ng, 2012: Theoretical and
  experimental study of particle trajectories for nonlinear water waves
  propagating on a sloping bottom. \textit{Philosophical Transactions of the
  Royal Society A: Mathematical, Physical and Engineering Sciences},
  \textbf{370~(1964)}, 1543--1571, \doi{10.1098/rsta.2011.0446}.

\bibitem[{Datawell(2019{\natexlab{a}})}]{DWR4manual}
Datawell, 2019{\natexlab{a}}: \textit{Datawell Waverider Manual: {DWR4}}. The
  Netherlands, Datawell BV.

\bibitem[{Datawell(2019{\natexlab{b}})}]{DWRMkIIImanual}
Datawell, 2019{\natexlab{b}}: \textit{Datawell Waverider Reference Manual:
  {DWR-MkIII, DWR-G, WR-SG}}. The Netherlands, Datawell BV.

\bibitem[{Datawell(n.d.)}]{DWRcomparison}
Datawell, n.d.: \textit{A comparative report on the {DWR MkIII} and {DWR4}
  data}. The Netherlands, Datawell BV.

\bibitem[{Didenkulova(2020)}]{DIDENKULOVA2020105076}
Didenkulova, E., 2020: Catalogue of rogue waves occurred in the world ocean
  from 2011 to 2018 reported by mass media sources. \textit{Ocean and Coastal
  Management}, \textbf{188}, 105\,076, \doi{10.1016/j.ocecoaman.2019.105076}.

\bibitem[{Dysthe et~al.(2008)Dysthe, Krogstad,, and Müller}]{Dysthe2008}
Dysthe, K., H.~E. Krogstad, and P.~Müller, 2008: Oceanic rogue waves.
  \textit{Annual Review of Fluid Mechanics}, \textbf{40~(1)}, 287--310,
  \doi{10.1146/annurev.fluid.40.111406.102203}.

\bibitem[{Fedele et~al.(2016)Fedele, Brennan, Ponce~de Le{\'o}n, Dudley,, and
  Dias}]{fedele2016real}
Fedele, F., J.~Brennan, S.~Ponce~de Le{\'o}n, J.~Dudley, and F.~Dias, 2016:
  Real world ocean rogue waves explained without the modulational instability.
  \textit{Scientific reports}, \textbf{6~(1)}, 1--11.

\bibitem[{Forristall(1984)}]{Forristall1984}
Forristall, G.~Z., 1984: The distribution of measured and simulated wave
  heights as a function of spectral shape. \textit{Journal of Geophysical
  Research: Oceans}, \textbf{89~(C6)}, 10\,547--10\,552,
  \doi{https://doi.org/10.1029/JC089iC06p10547}.

\bibitem[{Forristall(2000)}]{Forristall2000}
Forristall, G.~Z., 2000: Wave crest distributions: {O}bservations and
  second-order theory. \textit{Journal of Physical Oceanography},
  \textbf{30~(8)}, 1931--1943,
  \doi{10.1175/1520-0485(2000)030<1931:WCDOAS>2.0.CO;2}.

\bibitem[{Gramstad et~al.(2013)Gramstad, Zeng, Trulsen,, and
  Pedersen}]{doi:10.1063/1.4847035}
Gramstad, O., H.~Zeng, K.~Trulsen, and G.~K. Pedersen, 2013: Freak waves in
  weakly nonlinear unidirectional wave trains over a sloping bottom in shallow
  water. \textit{Physics of Fluids}, \textbf{25~(12)}, 122\,103,
  \doi{10.1063/1.4847035}.

\bibitem[{Haver(2004)}]{haver2004possible}
Haver, S., 2004: A possible freak wave event measured at the {D}raupner
  {J}acket {J}anuary 1 1995. \textit{Rogue waves}, Vol. 2004, 1--8.

\bibitem[{Haver and Andersen(2000)Haver, and Andersen}]{haver2000rare}
Haver, S., and O.~J. Andersen, 2000: Freak waves: {R}are realizations of a
  typical population or typical realizations of a rare population?
  \textit{International Offshore and Polar Engineering Conference}, Seattle,
  International Society of Offshore and Polar Engineers.

\bibitem[{Herbers and Janssen(2016)Herbers, and
  Janssen}]{10.1175/JPO-D-15-0129.1}
Herbers, T. H.~C., and T.~T. Janssen, 2016: Lagrangian surface wave motion and
  {S}tokes drift fluctuations. \textit{Journal of Physical Oceanography},
  \textbf{46~(4)}, 1009--1021, \doi{10.1175/JPO-D-15-0129.1}.

\bibitem[{Herterich and Dias(2019)Herterich, and Dias}]{herterich_dias_2019}
Herterich, J.~G., and F.~Dias, 2019: Extreme long waves over a varying
  bathymetry. \textit{Journal of Fluid Mechanics}, \textbf{878}, 481--501,
  \doi{10.1017/jfm.2019.618}.

\bibitem[{Hunt(2003)}]{hunt2003extreme}
Hunt, A., 2003: Extreme waves, overtopping and flooding at sea defences. Ph.D.
  thesis, University of Oxford.

\bibitem[{Jonathan and Taylor(1997)Jonathan, and Taylor}]{10.1115/1.2829043}
Jonathan, P., and P.~H. Taylor, 1997: On irregular, nonlinear waves in a spread
  sea. \textit{Journal of Offshore Mechanics and Arctic Engineering},
  \textbf{119~(1)}, 37--41, \doi{10.1115/1.2829043}.

\bibitem[{Judge et~al.(2019)Judge, Hunt-Raby, Orszaghova, Taylor,, and
  Borthwick}]{JUDGE2019103531}
Judge, F.~M., A.~C. Hunt-Raby, J.~Orszaghova, P.~H. Taylor, and A.~G.
  Borthwick, 2019: Multi-directional focused wave group interactions with a
  plane beach. \textit{Coastal Engineering}, \textbf{152}, 103\,531,
  \doi{10.1016/j.coastaleng.2019.103531}.

\bibitem[{Lenain et~al.(2019)Lenain, Pizzo,, and
  Melville}]{lenain_pizzo_melville_2019}
Lenain, L., N.~Pizzo, and W.~K. Melville, 2019: Laboratory studies of
  {L}agrangian transport by breaking surface waves. \textit{Journal of Fluid
  Mechanics}, \textbf{876}, R1, \doi{10.1017/jfm.2019.544}.

\bibitem[{Lindgren(1970)}]{10.2307/2240325}
Lindgren, G., 1970: Some properties of a normal process near a local maximum.
  \textit{The Annals of Mathematical Statistics}, \textbf{41~(6)}, 1870--1883.

\bibitem[{Lindgren and Rychlik(1991)Lindgren, and Rychlik}]{10.2307/1403443}
Lindgren, G., and I.~Rychlik, 1991: Slepian models and regression
  approximations in crossing and extreme value theory. \textit{International
  Statistical Review / Revue Internationale de Statistique}, \textbf{59~(2)},
  195--225.

\bibitem[{Longuet-Higgins(1980)}]{Longuet-Higgins1980}
Longuet-Higgins, M.~S., 1980: On the distribution of the heights of sea waves:
  Some effects of nonlinearity and finite band width. \textit{Journal of
  Geophysical Research: Oceans}, \textbf{85~(C3)}, 1519--1523,
  \doi{https://doi.org/10.1029/JC085iC03p01519}.

\bibitem[{Lygre and Krogstad(1986)Lygre, and Krogstad}]{LygreKrogstad1986}
Lygre, A., and H.~E. Krogstad, 1986: Maximum entropy estimation of the
  directional distribution in ocean wave spectra. \textit{Journal of Physical
  Oceanography}, \textbf{16~(12)}, 2052--2060,
  \doi{10.1175/1520-0485(1986)016<2052:MEEOTD>2.0.CO;2}.

\bibitem[{Mackay(2009)}]{soton79448}
Mackay, E. B.~L., 2009: Wave energy resource assessment. Ph.D. thesis,
  University of Southampton,
  \urlprefix\url{https://eprints.soton.ac.uk/79448/}.

\bibitem[{Madsen et~al.(2002)Madsen, Bingham,, and Liu}]{madsen2002}
Madsen, P.~A., H.~B. Bingham, and H.~Liu, 2002: A new {B}oussinesq method for
  fully nonlinear waves from shallow to deep water. \textit{Journal of Fluid
  Mechanics}, \textbf{462}, 1--30, \doi{10.1017/S0022112002008467}.

\bibitem[{Magnusson and Donelan(2013)Magnusson, and
  Donelan}]{Magnusson2013Andrea}
Magnusson, A.~K., and M.~A. Donelan, 2013: The {A}ndrea wave characteristics of
  a measured {N}orth {S}ea rogue wave. \textit{Journal of Offshore Mechanics
  and Arctic Engineering}, \textbf{135~(3)}, \doi{10.1115/1.4023800}.

\bibitem[{McAllister et~al.(2019)McAllister, Draycott, Adcock, Taylor,, and
  van~den Bremer}]{mcallister_draycott_adcock_taylor_van_den_bremer_2019}
McAllister, M.~L., S.~Draycott, T.~A.~A. Adcock, P.~H. Taylor, and T.~S.
  van~den Bremer, 2019: Laboratory recreation of the {Draupner} wave and the
  role of breaking in crossing seas. \textit{Journal of Fluid Mechanics},
  \textbf{860}, 767--786, \doi{10.1017/jfm.2018.886}.

\bibitem[{McAllister and van~den Bremer(2020)McAllister, and van~den
  Bremer}]{10.1175/JPO-D-19-0228.1}
McAllister, M.~L., and T.~S. van~den Bremer, 2020: Experimental study of the
  statistical properties of directionally spread ocean waves measured by buoys.
  \textit{Journal of Physical Oceanography}, \textbf{50~(2)}, 399--414,
  \doi{10.1175/JPO-D-19-0228.1}.

\bibitem[{Mori et~al.(2002)Mori, Liu,, and Yasuda}]{MORI20021399}
Mori, N., P.~C. Liu, and T.~Yasuda, 2002: Analysis of freak wave measurements
  in the {Sea of Japan}. \textit{Ocean Engineering}, \textbf{29~(11)},
  1399--1414, \doi{10.1016/S0029-8018(01)00073-7}.

\bibitem[{Nelson(1987)}]{nelson1987design}
Nelson, R.~C., 1987: Design wave heights on very mild slopes: An experimental
  study. \textit{Civil Engrg. Trans., Inst. Engrs. Aust.}, \textbf{29~(3)},
  157--161.

\bibitem[{O'Brien et~al.(2018)O'Brien, Renzi, Dudley, Clancy,, and
  Dias}]{nhess-18-729-2018}
O'Brien, L., E.~Renzi, J.~M. Dudley, C.~Clancy, and F.~Dias, 2018: Catalogue of
  extreme wave events in {Ireland}: {R}evised and updated for 14\,680 {BP} to
  2017. \textit{Natural Hazards and Earth System Sciences}, \textbf{18~(3)},
  729--758, \doi{10.5194/nhess-18-729-2018}.

\bibitem[{{Olhede} and {Walden}(2002){Olhede}, and {Walden}}]{1041025}
{Olhede}, S.~C., and A.~T. {Walden}, 2002: Generalized {M}orse wavelets.
  \textit{IEEE Transactions on Signal Processing}, \textbf{50~(11)},
  2661--2670, \doi{10.1109/TSP.2002.804066}.

\bibitem[{Orszaghova et~al.(2012)Orszaghova, Borthwick,, and
  Taylor}]{ORSZAGHOVA2012328}
Orszaghova, J., A.~G. Borthwick, and P.~H. Taylor, 2012: From the paddle to the
  beach – a {Boussinesq} shallow water numerical wave tank based on {Madsen
  and Sørensen’s} equations. \textit{Journal of Computational Physics},
  \textbf{231~(2)}, 328--344, \doi{10.1016/j.jcp.2011.08.028}.

\bibitem[{Orszaghova et~al.(2014)Orszaghova, Taylor, Borthwick,, and
  Raby}]{ORSZAGHOVA201463}
Orszaghova, J., P.~H. Taylor, A.~G. Borthwick, and A.~C. Raby, 2014: Importance
  of second-order wave generation for focused wave group run-up and
  overtopping. \textit{Coastal Engineering}, \textbf{94}, 63--79,
  \doi{doi.org/10.1016/j.coastaleng.2014.08.007}.

\bibitem[{Srokosz and Longuet-Higgins(1986)Srokosz, and
  Longuet-Higgins}]{srokosz_longuet-higgins_1986}
Srokosz, M.~A., and M.~S. Longuet-Higgins, 1986: On the skewness of sea-surface
  elevation. \textit{Journal of Fluid Mechanics}, \textbf{164}, 487--497,
  \doi{10.1017/S0022112086002653}.

\bibitem[{Tayfun and Fedele(2007)Tayfun, and Fedele}]{TAYFUN20071631}
Tayfun, M.~A., and F.~Fedele, 2007: Wave-height distributions and nonlinear
  effects. \textit{Ocean Engineering}, \textbf{34~(11)}, 1631--1649,
  \doi{10.1016/j.oceaneng.2006.11.006}.

\bibitem[{Tromans et~al.(1991)Tromans, Anaturk,, and
  Hagemeijer}]{tromans1991new}
Tromans, P.~S., A.~R. Anaturk, and P.~Hagemeijer, 1991: A new model for the
  kinematics of large ocean waves - application as a design wave. \textit{The
  First International Offshore and Polar Engineering Conference}, Edinburgh,
  International Society of Offshore and Polar Engineers.

\bibitem[{Trulsen et~al.(2012)Trulsen, Zeng,, and
  Gramstad}]{doi:10.1063/1.4748346}
Trulsen, K., H.~Zeng, and O.~Gramstad, 2012: Laboratory evidence of freak waves
  provoked by non-uniform bathymetry. \textit{Physics of Fluids},
  \textbf{24~(9)}, 097\,101, \doi{10.1063/1.4748346}.

\bibitem[{Tucker(1993)}]{TUCKER1993459}
Tucker, M., 1993: Recommended standard for wave data sampling and
  near-real-time processing. \textit{Ocean Engineering}, \textbf{20~(5)},
  459--474, \doi{10.1016/0029-8018(93)90015-A}.

\bibitem[{Underwood(2013)}]{Underwood2013}
Underwood, R., 2013: Lucky escapes in {T}orndirrup {N}ational {P}ark.
  \textit{Landscope}, \textbf{29~(1)}, 53--56.

\bibitem[{Voermans et~al.(2021)Voermans, Babanin, Kirezci, Carvalho, Santini,
  Pavani,, and Pezzi}]{Voermans2021}
Voermans, J.~J., A.~V. Babanin, C.~Kirezci, J.~T. Carvalho, M.~F. Santini,
  B.~F. Pavani, and L.~P. Pezzi, 2021: Wave anomaly detection in wave
  measurements. \textit{Journal of Atmospheric and Oceanic Technology},
  \textbf{38~(3)}, 525--536, \doi{10.1175/JTECH-D-20-0090.1}.

\bibitem[{Whittaker et~al.(2016)Whittaker, Raby, Fitzgerald,, and
  Taylor}]{WHITTAKER2016253}
Whittaker, C.~N., A.~C. Raby, C.~J. Fitzgerald, and P.~H. Taylor, 2016: The
  average shape of large waves in the coastal zone. \textit{Coastal
  Engineering}, \textbf{114}, 253--264, \doi{10.1016/j.coastaleng.2016.04.009}.

\bibitem[{Zhang et~al.(2019)Zhang, Benoit, Kimmoun, Chabchoub,, and
  Hsu}]{zhang2019statistics}
Zhang, J., M.~Benoit, O.~Kimmoun, A.~Chabchoub, and H.-C. Hsu, 2019: Statistics
  of extreme waves in coastal waters: {L}arge scale experiments and advanced
  numerical simulations. \textit{Fluids}, \textbf{4~(2)}, 99,
  \doi{10.3390/fluids4020099}.

\end{thebibliography}

\end{document}